\documentclass[a4paper,3p,twocolumn,preprint]{elsarticle}

\journal{Physica D}

\usepackage{amsmath}
\usepackage{amssymb}
\usepackage{graphicx}
\usepackage{bm}
\usepackage{epstopdf}

\newcommand{\la}{\lambda}

\newcommand{\al}{\alpha}
\newcommand{\om}{\omega}

\newcommand{\prt}{\partial}
\newcommand{\ox}{\overline{x}}
\newcommand{\oq}{\overline{q}}

\begin{document}

\begin{frontmatter}

\title{
Asymptotic integrability and its consequences}

\author{A.~M.~Kamchatnov}
\ead{kamch@isan.troitsk.ru}

\address{
Institute of Spectroscopy, Russian Academy of Sciences, Troitsk,
Moscow, 142190, Russia }
\address{
Skolkovo Institute of Science and Technology, Skolkovo, Moscow, 143026, Russia}

\date{\today}

\begin{abstract}
We give a brief review of the concept of asymptotic integrability, which means that the Hamilton 
equations for the propagation of short-wavelength packets along a smooth, large-scale background 
wave have an integral independent of the initial conditions. The existence of such an integral 
leads to a number of important consequences, which include, besides the direct application to 
the packets propagation problems, Hamiltonian theory of narrow solitons motion and generalized 
Bohr-Sommerfeld rule for parameters of solitons produced from an intensive initial pulse. 
We show that in the case of systems with two wave variables and exact fulfillment of the 
asymptotic integrability condition, the `quantization' of mechanical systems, associated with
the additional integrals, yields the Lax 
pairs for a number of typical completely integrable equations, and this sheds new light on 
the origin of the complete integrability in nonlinear wave physics.
\end{abstract}



\end{frontmatter}


{\it Dedicated to the memory of V. E. Zakharov}

\section{Introduction}

The discovery of the inverse scattering transform (IST) method for integration of the Korteweg-de Vries 
(KdV) equation by M. D. Kruskal and co-authors \cite{ggkm-67} ushered in a new epoch in nonlinear 
physics. Its more general formulation given by P. D. Lax \cite{lax-68} suggested possible generality 
of the IST method, and this possibility was realized in the papers \cite{zs-71,zs-73} by 
V. E. Zakharov and A. B. Shabat, where the IST method was extended to the nonlinear Schrödinger (NLS) 
equation for both signs of the nonlinear term. This remarkable achievement resulted in a flood of 
discoveries of a number of other nonlinear wave equations integrable by the IST method. At last, 
V. E. Zakharov and L. D. Faddeev \cite{zf-71} and S. C. Gardner \cite{gardner-71} showed that the IST 
method is nothing but the transition from physical variables to the action-angle variables generalized to 
systems with an infinite number of degrees of freedom, and this Hamiltonian system is completely 
integrable in the Liouville-Arnold sense \cite{liouville,arnold}. This demonstration related the IST 
method with classical Hamiltonian mechanics and stimulated fast development of modern nonlinear 
mathematical physics (see, e.g., \cite{nmpz-80,as-81,newell-85,dickey,ft-07} and references therein). 
However, the relationship between physical properties of wave equations and their complete 
integrability still seems not clear enough.

As was recently noticed \cite{kamch-24,kamch-25}, if we confine ourselves to integrability of 
nonlinear wave equations in a restricted sense, formulated as the condition of integrability of 
the Hamilton equations that govern the propagation of short-wavelength packets along a large-scale 
(hydrodynamic) background wave, then the fulfilment of this {\it asymptotic integrability} condition 
means the existence of the integral of the Hamilton equations, that is the carrier wave number $k$ 
becomes a function of local background wave variables. The very existence of such an integral leads 
to a number of important consequences.

First of all, the expression for the carrier wave number greatly simplifies solutions of problems 
related to the propagation of linear wave packets along large-scale background waves 
\cite{ceh-18,sk-23}. Next, if such a packet corresponds to a small-amplitude edge of a dispersive 
shock wave in the Gurevich-Pitaevskii theory of these shock waves \cite{gp-73}, based on the Whitham 
theory of modulations \cite{whitham-65,whitham} (see also \cite{eh-16,kamch-21}), then this expression 
allows one to find a path of this edge during the evolution of the shock \cite{el-05,kamch-19}. 
Then, as was noticed by A.~V.~Gurevich and L.~P.~Pitaevskii \cite{gp-87}, their theory yields the expression 
for the speed of entering of oscillations into the dispersive shock region. If the initial wave pulse 
is localized in space, then these oscillations transform eventually into a train of separate solitons, 
and, consequently, one can calculate the final number of solitons at asymptotically large time 
\cite{kamch-21,kamch-20} for a given initial profile. Natural extension of this theory \cite{kamch-23} 
yields the generalized Bohr-Sommerfeld quantization rule for finding the parameters of these asymptotic 
solitons, which generalizes earlier theory \cite{egs-08} developed 
for unidirectional initial pulses.

Besides that, the above-mentioned integral of Hamilton equations for a packet's motion can be transformed 
to a similar expression relating the inverse half-width $\kappa$ of narrow solitons with the local 
values of large-scale background wave variables. This is achieved by means of Stokes' reasoning 
\cite{stokes} who noticed that the exponentially small soliton tails $\propto\exp[\pm\kappa(x-Vt)]$ 
and the linear harmonic waves $\propto\exp[i(kx-\om t)]$ obey the same linearized equations and, 
hence, the phase velocity $V=\om(k)/k$ converts to the soliton velocity $V=\om(i\kappa)/(i\kappa)$ 
by replacing $k\to i\kappa$. The existence of such an integral for the dynamics of a soliton 
propagating along a non-uniform and non-stationary background wave allows one to develop the Hamilton 
theory of soliton dynamics  \cite{kamch-25} (see also \cite{ik-22,ks-23,kamch-24b,ks-24,kamch-24c}). 

At last, the additional integrals are related to the quasiclassical limit of Lax pairs in
the Ablowitz-Kaup-Newell-Segur (AKNS) scheme \cite{akns-74}, and these limiting formulas are expressed in
terms of the dispersion relation for linear harmonic waves considered in the framework of the
corresponding nonlinear wave equation \cite{ks-24a}.  We show that different methods of quantization of 
mechanical systems associated with the additional integrals lead to the Lax pairs for a number of 
completely integrable equations, in particular, for the completely
integrable equations with `energy-dependent potentials' \cite{jm-76,pavlov-14} and for NLS \cite{zs-71,zs-73}
and derivative NLS \cite{kn-78} equations.

It is remarkable that even if the condition of asymptotic integrability is
only fulfilled for some classes of background large-scale waves or if it is fulfilled only approximately,
then it results in the existence of the exact or approximate integral of Hamilton's equations for the 
packet's motion, so that the completely integrable equations share some their properties with not
completely integrable ones.

In this paper, we will present a unified and consistent exposition of the asymptotic integrability 
approach to the study of nonlinear wave equations.

\section{Basic equations}

At first, we have to discuss the basic dispersive and nonlinear properties that characterize
the physical system under consideration. We assume that it is described by two variables $\rho$ 
and $u$ which we will call for definiteness as `density' and `flow velocity', correspondingly. 
They obey the system 
\begin{equation}\label{eq1}
\begin{split}
  &\rho_t+F(\rho,u,\rho_x,u_x,\ldots)=0,\\
  &u_t+G(\rho,u,\rho_x,u_x,\ldots)=0,
  \end{split}
\end{equation}
where dots denote higher order $x$ derivatives. As a rule, such a system has solutions 
$\rho=\rho_0=\mathrm{const}$, $u=u_0=\mathrm{const}$, which describe uniform states of the
physical system with $\rho_0,u_0$ belonging to some intervals of values (usually, $\rho_0>0$). 
If such a state is slightly disturbed, then waves with infinitely small amplitudes
$\rho'=\rho-\rho_0, u'=u-u_0$ propagate along it, and their dynamics is governed by Eqs.~(\ref{eq1})
linearized with respect to $\rho',u'$. This linearized system has harmonic wave solutions
\begin{equation}\label{eq2}
  \rho',u'\propto\exp[i(kx-\om t)],
\end{equation}
if $\om$ is related with $k$ by the dispersion relation
\begin{equation}\label{eq3}
  \om=\om_{\pm}(k,\rho_0,u_0),
\end{equation}
where $\pm$-signs denote the two branches of linear modes. The dispersion relation (\ref{eq3})
characterizes the basic dispersive properties of our physical system.

In the opposite dispersionless limit, we neglect dispersive properties, that is we omit all the
higher-order derivatives or products of derivatives in Eqs.~(\ref{eq1}), but we do not impose
any restrictions on the amplitudes of the variables $\rho,u$. Then we arrive at the `hydrodynamic' 
limit of Eqs.~(\ref{eq1}). We assume that the resulting equations can be written in a quasilinear
form
\begin{equation}\label{eq4}
  \left(
          \begin{array}{c}
            \rho \\
            u \\
          \end{array}
        \right)_t+ \mathbb{A}
        \left(
          \begin{array}{c}
            \rho \\
            u \\
          \end{array}
        \right)_x=0,\quad
        \mathbb{A}=\left(
                     \begin{array}{cc}
                       a_{11} & a_{12} \\
                       a_{21} & a_{22} \\
                     \end{array}
                   \right),
\end{equation}
where $a_{ij}=a_{ij}(\rho,u)$. This system characterizes the nonlinear properties of our
physical system.

The theory acquires a more universal form, if we transform the system (\ref{eq4}) from
the physical variables $\rho,u$ to more convenient Riemann invariants $r_{\pm}=r_{\pm}(\rho,u)$,
so that the dispersionless equations take a diagonal form
\begin{equation}\label{eq5}
  \frac{\prt r_+}{\prt t}+v_+\frac{\prt r_+}{\prt x}=0,\quad
  \frac{\prt r_-}{\prt t}+v_-\frac{\prt r_-}{\prt x}=0.
\end{equation}
This can always be done for systems with two wave variables (see, e.g., \cite{ry-83,kamch-book}).
Then the physical variables $\rho,u$ become some functions $\rho=\rho(r_+,r_-),u=u(r_+,r_-)$
of the Riemann invariants, 
and their substitution into Eq.~(\ref{eq3}) yields the dispersion relation in the form
\begin{equation}\label{eq6}
  \om=\om_{\pm}(k,r_+,r_-).
\end{equation}
The velocities $v_{\pm}$ in Eqs.~(\ref{eq5}) are just the long wavelength limits
of the phase velocities of the harmonic waves, i.e.
\begin{equation}\label{eq7}
  v_{\pm}=\lim_{k\to0}\frac{\om_{\pm}(k,r_+,r_-)}{k}.
\end{equation}

Strictly speaking, the dispersion relation (\ref{eq6}) is defined for a uniform state with 
constant values of the Riemann invariants $r_{\pm}$. However, if the characteristic size $l$
along which $r_{\pm}$ considerably change is much greater than the wavelength $2\pi/k$ of
a harmonic wave, then in such a wave the frequency $\om$ and the wave number $k$ are still related 
by Eq.~(\ref{eq6}) with a good accuracy, where $r_+$ and $r_-$ are considered as slowly
changing parameters. In this case of large difference between the two scales $l$ and $2\pi/k$,
we can build a wave packet with width $\Delta$ which satisfies the conditions
\begin{equation}\label{eq8}
  l\gg\Delta\gg2\pi/k.
\end{equation}
This means that we can define a coordinate $x(t)$ of the packet with the accuracy $\Delta$ and
the wave number $k(t)$ with the accuracy $\sim2\pi/\Delta\ll k$, and then propagation of such a
packet along a non-uniform and time-dependent large-scale background wave $r_{\pm}=r_{\pm}(x,t)$
obeys the Hamilton equations \cite{kamch-book,LL-2,ko-90}
\begin{equation}\label{eq9}
  \frac{dx}{dt}=\frac{\prt\om}{\prt k}, \qquad \frac{dk}{dt}=-\frac{\prt\om}{\prt x},
\end{equation}
where $\om(k)$ corresponds to a chosen branch of the small-amplitude wave propagation.
Here the first equation corresponds to the well-known
definition of the group velocity
\begin{equation}\label{eq10}
  v_g=\frac{\prt\om}{\prt k},
\end{equation}
and the second equation describes refraction, i.e., change of the wavelength in a
non-uniform medium.

The convenience of the Riemann invariants for wave physics is related with the fact that one of
them is constant in the case of a unidirectional wave propagation, and such waves are called `simple waves'.
Let, e.g., $r_-$ be constant, then the system (\ref{eq5}) reduces to a single equation for the
$r_+$-invariant. At the same time, $\rho$ and $u$ are related with each other by the condition
$r_-(\rho,u)=\mathrm{const}$. Consequently, one of the variables, say $\rho$, can be eliminated,
and we arrive at the so-called Hopf equation
\begin{equation}\label{eq11}
  u_t+V_0(u)u_x=0.
\end{equation}
If a short wavelength propagates along such a simple wave, then it satisfies the dispersion relation
\begin{equation}\label{eq12}
  \om=\om(k,u).
\end{equation}
Motion of wave packets still obeys the Hamilton equations (\ref{eq9}) with $\om$ given by Eq.~(\ref{eq12}).

The solution $u=u(x,t)$ of Eq.~(\ref{eq11}) for the large-scale background wave can easily be found
by the method of characteristics (see, e.g., \cite{whitham,ry-83,kamch-book}) and it is given in an
implicit form by the formula
\begin{equation}\label{eq13}
  x-V_0(u)t=\ox(u),
\end{equation}
where $\ox(u)$ is a function inverse to the initial distribution $u=u_0(x)\equiv u(x,0)$ of the
variable $u$. A similar solution can often be written in a more complicated form for the system 
(\ref{eq5}) in the framework of the hodograph method (see, e.g., \cite{whitham,ry-83,kamch-book}),
but we will not use this solution here.
 
To illustrate our definitions by a concrete example, let us consider the generalized NLS
(or Gross-Pitaevskii) equation
\begin{equation}\label{eq14}
i\psi_t+\frac{1}{2}\psi_{xx} - f(|\psi|^2)\psi = 0,
\end{equation}
where it is assumed that the function $f(|\psi|^2)$ is positive and it increases monotonously with growth 
of its argument. It can be cast to the form of Eqs.~(\ref{eq1}) by means of the Madelung substitution
\begin{equation}\label{eq15}
\psi(x,t) = \sqrt{\rho(x,t)}\exp\left(i\int^x u(x',t)dx'\right),
\end{equation}
so we get
\begin{equation}\label{eq16}
\begin{split}
&\rho_t + (\rho u)_x = 0, \\
&u_t + uu_x + \frac{c^2}{\rho} \rho_x + \bigg( \frac{\rho_x^2}{8\rho^2}
- \frac{\rho_{xx}}{4\rho} \bigg)_x = 0,
\end{split}
\end{equation}
where
\begin{equation}\label{eq17}
  c^2=\rho f'(\rho).
\end{equation}
Linearization of this system leads to the Bogoliubov dispersion relation
\begin{equation}\label{eq18}
\omega = k\left( u \pm \sqrt{c^2+\frac{k^2}{4}} \right),
\end{equation}
hence $c=c(\rho)$ has the meaning of the sound velocity of long waves propagating along the
medium with constant density $\rho$ and zero flow velocity $u$. In the dispersionless
limit the system (\ref{eq16}) reduces to the standard gas dynamics equations
\begin{equation}\label{eq19}
\begin{split}
\rho_t + (\rho u)_x = 0, \quad
u_t + uu_x + \frac{c^2}{\rho} \rho_x = 0.
\end{split}
\end{equation} 
The velocities $v_{\pm}$ in Eqs.~(\ref{eq5}) are equal to $v_{\pm}=u\pm c$, that is they
correspond to the long wavelength sound waves propagating downstream or upstream
the flow $u$. It is convenient to define the Riemann invariants by the formulas
\begin{equation}\label{eq20}
  r_{\pm}(\rho,u)=\frac{u}{2}\pm\frac12\int_0^{\rho}\frac{cd\rho}{\rho}
\end{equation}
or
\begin{equation}\label{eq21}
\begin{split}
  &r_{\pm}(c,u)=\frac{u}{2}\pm \sigma(c),\\
  &\sigma(c)=\frac12\int_0^c\frac{c}{\rho(c)}\frac{d\rho}{dc}dc,
  \end{split}
\end{equation}
where $\rho=\rho(c)$ is the inverse of the function $c=c(\rho)$ defined by Eq.~(\ref{eq17}).
The sound velocity $c$ can replace the density $\rho$ as one of the background wave variables. 
In an important particular case with $f(\rho)=\rho$ we have $c^2=\rho$ and
\begin{equation}\label{eq22}
  r_{\pm}=\frac{u}{2}\pm\sqrt{\rho}=\frac{u}{2}\pm c.
\end{equation}

Now we can formulate the condition of asymptotic integrability.

\section{Asymptotic integrability}

The asymptotic integrability condition is formulated especially simple, if a short
wavelength packet propagates along a simple wave background $u=u(x,t)$ obeying the Hopf
equation (\ref{eq11}). In this case with $\om=\om(k,u)$, the wave number $k$ changes
along the packet's path according to the equation
$$
\frac{dk}{dt}=-\frac{\prt \om}{\prt u}\frac{\prt u}{\prt x}.
$$
But we have
$$
\frac{du}{dt}=\frac{\prt u}{\prt t}+\frac{\prt u}{\prt x}\frac{dx}{dt}
=-\left(V_0-v_g\right)\frac{\prt u}{\prt x},
$$
where we used Eqs.~(\ref{eq10}) and (\ref{eq11}). Hence, the ratio of these two expressions
gives the equation
\begin{equation}\label{eq23}
  \frac{dk}{du}=\frac{\prt\om/\prt u}{V_0-v_g}.
\end{equation}
Here the right-hand side is a function of $k$ and $u$, so this differential equation
yields a solution
\begin{equation}\label{eq24}
  k=k(u,q),
\end{equation}
where $q$ is an integration constant. The expression (\ref{eq24}) can be considered as
an integral of the Hamilton equations (\ref{eq9}), so propagation of wave packets along
simple waves is an asymptotically integrable problem and the condition of asymptotic
integrability does not impose any restrictions on the form of the dispersion relation
$\om(k,u)$.

In fact, Eq.~(\ref{eq23}) was first derived by El \cite{el-05} in the framework of
Gurevich-Pitaevskii theory of dispersive shock waves as a consequence of the `number of
waves' conservation law
\begin{equation}\label{eq25}
  k_t+\om_x=0
\end{equation}
in the Whitham theory of modulations of nonlinear waves \cite{whitham-65,whitham}
applied to the problem of propagation of a small-amplitude edge of a shock. Relation of
Eq.~(\ref{eq23}) to the general optical-mechanical analogy was explained in Ref.~\cite{kamch-20}.

As a simple example, let us consider the generalized KdV equation \cite{el-05}
\begin{equation}\label{eq26}
  u_t+V_0(u)u_x+u_{xxx}=0.
\end{equation}
Its linearization yields the dispersion relation
\begin{equation}\label{eq27}
  \om=V_0(u)k-k^3,
\end{equation}
so $v_g=V_0-3k^2$ and Eq.~(\ref{eq23}) gives the solution
\begin{equation}\label{eq28}
  k^2=\frac23(V_0(u)-q),
\end{equation}
where $q$ is an integration constant.

As another important example, we consider propagation of a wave packet along a simple wave in 
the theory of the generalized NLS equation (\ref{eq15}) \cite{hoefer-14}. We assume that
the Riemann invariant $r_-$ is constant (see Eqs. (\ref{eq21})),
\begin{equation}\label{eq29}
  r_-=u/2-\sigma(c)=-\sigma(c^*),
\end{equation}
where $c^*$ denotes the sound velocity at the point with vanishing flow velocity $u$.
Consequently,
\begin{equation}\label{eq30}
  u(c)=2[\sigma(c)-\sigma(c^*)],\quad r_+=2\sigma(c)-\sigma(c^*),
\end{equation}
and the first Eq.~(\ref{eq5}) can be written in the form
\begin{equation}\label{eq31}
  c_t+V_0(c)c_x=0,
\end{equation}
where
\begin{equation}\label{eq32}
 V_0(c)=u(c)+c=2[\sigma(c)-\sigma(c^*)]+c.
\end{equation}
It is convenient to write the dispersion relation (\ref{eq18}) as
\begin{equation}\label{eq33}
\begin{split}
  \om(k)&=k[u+c\al(c)]\\
  &=k\left\{2[\sigma(c)-\sigma(c_R)]+c\al(c)\right\},
  \end{split}
\end{equation}
where
\begin{equation}\label{eq34}
  \al(c)=\sqrt{1+\frac{k^2}{4c^2}},\quad k(c)=2c\sqrt{\al^2(c)-1}.
\end{equation}
Then Eq.~(\ref{eq23}) with $c$ as an independent variable can be transformed to
\begin{equation}\label{eq35}
  \frac{d\al}{dc}=-\frac{(\al+1)(2\sigma'+2\al-1)}{c(2\al+1)}.
\end{equation}
This equation can be solved in a closed form for the case of (see Ref.~\cite{hoefer-14})
\begin{equation}\label{eq36}
  f(\rho)=\frac1p\rho^p,\quad c=\rho^{p/2},\quad (p=\mathrm{const}),
\end{equation}
when $\sigma'(c)=1/p$, $\sigma(c)=c/p$, so we get ($p\neq 2/3$)
\begin{equation}\label{eq37}
\begin{split}
   c(\al)=q\left(\frac2{1+\al}\right)^{\frac{p}{3p-2}}\left(\frac{2+p}{2-p+2p\al}\right)
  ^{\frac{2(p-1)}{3p-2}},
  \end{split}
\end{equation}
where $q$ is an integration constant. In case of the completely integrable NLS
equation ($p=1$) we have
\begin{equation}\label{eq38}
  \al(c)=\frac{2q}{c}-1
\end{equation}
and substitution of this expression into the second Eq.~(\ref{eq34}) yields
\begin{equation}\label{eq39}
  k^2=16q(q-c).
\end{equation}
Explicit expressions for $\al(c)$ can also be obtained for the cases $p=2$ and $p=1/2$,
so we get, respectively,
\begin{equation}\label{eq40}
  \al(c)=\frac12\left(\sqrt{1+8\left(\frac{q}c\right)^2}-1\right),
\end{equation}
if $p=2$, and
\begin{equation}\label{eq41}
  \al(c)=\frac1{16}\left(5\sqrt{\frac{q}c\left(25\frac{q}c-16\right)}+25\frac{q}c-24\right),
\end{equation}
if $p=1/2$.

Other examples of solutions of Eq.~(\ref{eq23}) can be found in 
Refs.~\cite{el-05,egs-06,egkkk-07,lh-13,ep-14,ckp-16,sh-17,mfweh-20}.

The situation considerably changes in case of the general background large-scale wave with 
both Riemann invariants changing with space and time variables. Now we suppose that the 
Hamilton equations (\ref{eq9}) have an integral $Q(k,r_+,r_-)=\mathrm{const}$ which does
not depend on a particular choice of the background wave solution $r_{\pm}=r_{\pm}(x,t)$
of Eqs.~(\ref{eq5}). This means that the wave number $k$ can be considered as a function
$k=k(r_+,r_-)$ of the local values $r_+$ and $r_-$ of the Riemann invariants at the instant
location of the wave packet. Let us consider two moments of time separated by a small
interval $dt$ and a small distance $dx=v_gdt$ between the corresponding locations of the
packet. Then the corresponding differences between the values of the Riemann invariants
are equal to
$$
dr_{\pm}=\frac{\prt r_{\pm}}{\prt x}dx+\frac{\prt r_{\pm}}{\prt t}dt=
\frac{\prt r_{\pm}}{\prt x}(v_g-v_{\pm})dt,
$$
and, hence,
$$
\frac{\prt r_{\pm}}{\prt x}dt=-\frac{dr_{\pm}}{v_{\pm}-v_g}.
$$
Consequently, the second Hamilton equation (\ref{eq9}) yields the following
expression for the change of the wave number 
\begin{equation}\nonumber
  \begin{split}
  dk=-\frac{\prt\om}{\prt x}dt&=-\left(\frac{\prt\om}{\prt r_+}\frac{\prt r_+}{\prt x}+
  \frac{\prt\om}{\prt r_-}\frac{\prt r_-}{\prt x}\right)dt\\
  &=\frac{\prt\om/\prt r_+}{v_+-v_g}dr_++\frac{\prt\om/\prt r_-}{v_--v_g}dr_-.
  \end{split}
\end{equation}
On the other hand, our supposition that $k$ is a function $k=k(r_+,r_-)$ gives
$$
dk=\frac{\prt k}{\prt r_+}dr_+ +\frac{\prt k}{\prt r_-}dr_-.
$$
Since $dr_+$ and $dr_-$ correspond to an arbitrary solution of Eqs.~(\ref{eq5}), the
differentials $dr_+$ and $dr_-$ are independent of each other, and comparison of the 
above two expressions for $dk$ yields the system of equations
\begin{equation}\label{eq42}
  \frac{\prt k}{\prt r_+}=\frac{\prt\om/\prt r_+}{v_+-v_g},\quad
  \frac{\prt k}{\prt r_-}=\frac{\prt\om/\prt r_-}{v_--v_g}.
\end{equation}
This system has a solution, if these derivatives commute,
\begin{equation}\label{eq43}
  \frac{\prt}{\prt r_-}\left(\frac{\prt k}{\prt r_+}\right)=
   \frac{\prt}{\prt r_+}\left(\frac{\prt k}{\prt r_-}\right).
\end{equation}
The right-hand sides of Eqs.~(\ref{eq42}) are expressed in terms of the dispersion relation
(\ref{eq6}) (see Eqs.~(\ref{eq7}) and (\ref{eq10})). Thus, in this case the asymptotic
integrability condition imposes strong restrictions on the dispersion relation. We will
present here two examples when the condition (\ref{eq43}) is exactly satisfied, so that
Eqs.~(\ref{eq42}) yield the corresponding expressions for the integral of the Hamilton
equations.

In the first example the dispersion relation reads
\begin{equation}\label{eq44}
  \om=k\left(r_++r_-\pm\frac12\sqrt{(r_+-r_-)^2+\sigma{k^2}}\right),
\end{equation}
where $\sigma=\pm1$. 
Then, assuming $r_+>r_-$, we obtain
\begin{equation}\label{eq45}
  v_{\pm}=r_++r_-\pm\frac12(r_+-r_-),
\end{equation}
and Eqs.~(\ref{eq42}) yield the solution
\begin{equation}\label{eq46}
   k^2=4\sigma(q-r_+)(q-r_-),
\end{equation}
where $q$ is an integration constant.

In the second example we have
\begin{equation}\label{eq47}
\begin{split}
  \om=&k\Bigg\{r_++r_-+\frac{\sigma k^2}8\\
  &\pm\sqrt{\left(\frac{r_++r_-}2+\frac{\sigma k^2}8\right)^2-r_+r_-}\Bigg\},
  \end{split}
\end{equation}
so that for $r_+>r_-$ the velocities $v_{\pm}$ are given again by Eqs.~(\ref{eq45})
and we get the expression
\begin{equation}\label{eq48}
  k^2=\frac{4\sigma}{q}(q-r_+)(q-r_-)
\end{equation}
for the integral with arbitrary value of an integration constant $q$. 

It is easy to see that for Riemann invariants (\ref{eq22}) the dispersion relation
(\ref{eq44}) with $\sigma=1$ coincides exactly with Eq.~(\ref{eq18}), that is the NLS
equation (\ref{eq14}) with $f(\rho)=\rho$ is asymptotically integrable. In this case,
asymptotic integrability coincides with the complete integrability of the NLS equation 
established by Zakharov and Shabat in Ref.~\cite{zs-73}. However, in problems on
propagation of short wavelength packets we may be interested in knowledge of an
approximate integral of the Hamilton equations which is correct with good enough
accuracy only in the limit of large values of $k$. We will consider here such a possibility 
for the generalized NLS equation (\ref{eq14}). In this case, it is convenient to use 
$c$ and $u$ as the background wave variables instead of $r_+$ and $r_-$. Then 
Eqs.~(\ref{eq42}) take the form \cite{sk-23}
\begin{equation}\label{eq49}
  \begin{split}
  & \frac{\prt k}{\prt c}=-\frac{c[(2+c\rho'/\rho)k^2+4(1+c\rho'/\rho)c^2]}{k(k^2+3c^2)},\\
  & \frac{\prt k}{\prt u}=-\frac{\sqrt{k^2+4c^2}[k^2+2(1+\rho/(c\rho'))c^2]}{k(k^2+3c^2)},
  \end{split}
\end{equation}
and the difference of the cross-derivatives equals to
\begin{equation}\label{eq50}
  \begin{split}
  \frac{\prt}{\prt c}\left(\frac{\prt k}{\prt u}\right)&-\frac{\prt}{\prt u}\left(\frac{\prt k}{\prt c}\right)
  =\frac{\sqrt{k^2+4c^2}}{k\rho\rho'(k^2+3c^2)^2}\\
  &\times\big[(k^2+6c^2)\rho^{\prime 2}(\rho'-2c)\\
  &+2(k^2+3c^2)\rho^2(c\rho^{\prime\prime}-\rho')\big],
  \end{split}
\end{equation}
where $\rho=\rho(c)$ is the inverse of the function $c=c(\rho)$ defined by Eq.~(\ref{eq17}).
It is easy to see that this expression only vanishes for $\rho=c^2$, that is for the 
completely integrable NLS equation with $f(\rho)=\rho$, so that the integral (\ref{eq46})
takes the form
\begin{equation}\label{eq51}
  k^2=(q-u)^2-4\rho,
\end{equation}
where we redefined the integration constant $q\to q/2$.
In a particular case of a simple wave background with $u=2(c-c^*)$ we return to the integral
(\ref{eq39}) with another definition of an integration constant.

In general case with $f(\rho)\neq\rho$ the exact solution of Eqs.~(\ref{eq49}) does not exist,
but the difference of the cross-derivatives (\ref{eq50}) is proportional to $\propto k^{-2}$
in the limit $k\to\infty$, so we can find an approximate solution of Eqs.~(\ref{eq49}). In the
limit of large $k$ these equations reduce to
\begin{equation}\label{eq52}
  \frac{\prt k^2}{\prt c^2}=-2\left(1+\frac{c^2}{\rho}\frac{d\rho}{d c^2}\right),\qquad
  \frac{\prt k}{\prt u}=-1.
\end{equation}
The second equation gives $k=q-u+F(c^2)$, where in the main approximation $k\approx q\gg|u|,c$
and $|F|\ll q$. Then the first equation gives at once
$$
F(c^2)=-\frac1q(f(\rho)+\rho f'(\rho))
$$
and we get the expression
\begin{equation}\label{eq53}
  k^2=(q-u)^2-2\left(f(\rho)+\rho f'(\rho)\right),\quad q\gg|u|,c,
\end{equation}
for an approximate integral of the packet's Hamilton equations.

The existence of such an integral, exact or approximate, leads to important consequences which will be
considered below.

\section{Propagation of short-wavelength packets}

The most immediate application of the obtained above integrals is to the problem of propagation
of short-wavelength packets.
If the solution $r_+=r_+(x,t), r_-=r_-(x,t)$ of Eqs.~(\ref{eq5}) is known, then substitution of
these functions into the integral $k=k(r_+,r_-,q)$ gives us the wave number as a function $k=k(x,t)$ 
of the space and time variables provided the constant $q$ is determined by the value $k_0$ of the 
carrier wave number corresponding to the moment $t=0$ of time when the packet enters in the region 
of the background large-scale wave. Then the packet's path can be found by solving the equation
\begin{equation}\label{eq54}
  \frac{dx}{dt}=v_g(k(x,t),r_+(x,t),r_-(x,t))
\end{equation}
with some initial condition $x=x_0$ at $t=0$. Usually, this can only be done numerically and
some examples can be found in Ref.~\cite{sk-23}. However, if the background wave is a simple wave, 
then a more complete theory can be developed \cite{kamch-19}.

Let the background wave be a simple wave whose evolution is governed by the Hopf equation (\ref{eq11}),
and the initial distribution be given by the function $u=u_0(x)$, so that the inverse function
$x=\ox(u)$ can also be considered as known as well as the solution (\ref{eq13}) of the Hopf equation.
We assume that we know the solution (\ref{eq24}) of Eq.~(\ref{eq23}) for our physical problem. Then
our packet propagates with the known group velocity $v_g(k(u),u)$ and this expression suggests that
it is convenient to look for the path of the packet in a parametric form $t=t(u),x=x(u)$. 
Differentiation of Eq.~(\ref{eq13}) with respect to $u$ and exclusion of 
$\frac{dx}{du}=\frac{dx}{dt}\frac{dt}{du}=v_g\frac{dt}{du}$ yields a linear differential equation 
\begin{equation}\label{eq55}
  \left[v_g(k(u),u)-V_0(u)\right]\frac{dt}{du}-V_0'(u)t=\ox'(u)
\end{equation}
for the function $t=t(u)$. It should be solved with the condition that $t=0$ at the initial value of
$u$ at the initial location of the packet. Substitution of the function $t=t(u)$ into Eq.~(\ref{eq13})
yields the function $x=x(u)$, so the packet's path is found. Let us illustrate this approach by two
simple examples.

\subsection{Propagation of a packet along a rarefaction wave}\label{sub4-1}

Let our system be described by the KdV equation
\begin{equation}\label{eq56}
  u_t+6uu_x+u_{xxx}=0.
\end{equation}
Its linearization gives the dispersion relation
\begin{equation}\label{eq57}
  \om(k,u)=6uk-k^3,\quad\text{so}\quad v_g(k,u)=6u-3k^2.
\end{equation}
We assume that the background wave is represented by a rarefaction wave solution
\begin{equation}\label{eq58}
  u(x,t)=\frac{x}{6(t+t_0)},\quad 6u_-(t+t_0)\leq x\leq 0,
\end{equation}
of the Hopf equation (\ref{eq11}) with $V_0(u)=6u$. It connects the two plateau regions $u=0$
for $x>0$ and $u=u_-<0$ for $x<6u_-(t+t_0)$. The left edge of the rarefaction wave propagates
to the left with velocity $6u_-$ along the left plateau.

\begin{figure}[th]
\centerline{\includegraphics[ width=8cm,clip]{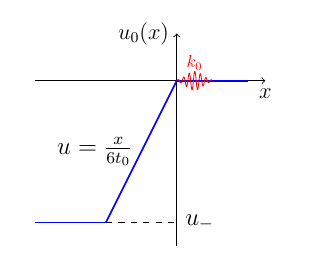}}
\caption{The initial distribution in the rarefaction wave (blue line) and a short-wavelength
packet entering into the rarefaction wave region (red line) at the initial moment of time.
}
\label{fig1}
\end{figure}

The packet with wave number $k_0$ enters at the moment of time $t=0$ into the rarefaction wave 
region from the right at $x=0$ (see Fig.~\ref{fig1}). Eq.~(\ref{eq23}) takes the form
\begin{equation}\label{eq59}
  \frac{dk}{du}=\frac2k,
\end{equation}
and its integration with the initial condition $k(0)=k_0$ yields
\begin{equation}\label{eq60}
  k^2=4u+k_0^2.
\end{equation}
Substitution of this expression into the formula for the group velocity (\ref{eq57}) gives
the equation
\begin{equation}\label{eq61}
  \frac{dx}{dt}=-6u-3k_0^2=-\frac{x}{t+t_0}-3k_0^2,
\end{equation}
and its integration with the initial condition $x(0)=0$ yields the formula for the packet's path,
\begin{equation}\label{eq62}
  x(t)=-\frac{3k_0^2}{2}\frac{t(t+2t_0)}{t+t_0}.
\end{equation}
If $|u_-|$ is large enough $(|u_-|>k_0^2/4)$, then at $t\to\infty$ the packet moves with the
constant velocity
\begin{equation}\label{eq63}
  v_g(\infty)=-\frac{3k_0^2}{2},
\end{equation}
that is slower than the left edge of the rarefaction wave ($3k_0^2/2<6|u_-|$). This means that
the packet remains forever inside the rarefaction wave region. It is easy to find from 
Eq.~(\ref{eq61}) that in this limit $x/(t+t_0)\to-3k_0^2/2$, $u\to-k_0^2/4$, and $k\to0$,
that is the packet disperses and our short-wavelength approximation loses its applicability.
If $|u_-|<k_0^2/4$, then the packet passes through the rarefaction wave and goes out of it at 
the left edge with the wave number
\begin{equation}\label{eq64}
  k_-=\sqrt{k_0^2+4u_-}<k_0.
\end{equation}
This problem was considered by another method in Ref.~\cite{ceh-18}.

\subsection{Propagation of a small-amplitude edge of a dispersive shock wave}

\begin{figure}[th]
\centerline{\includegraphics[ width=8cm,clip]{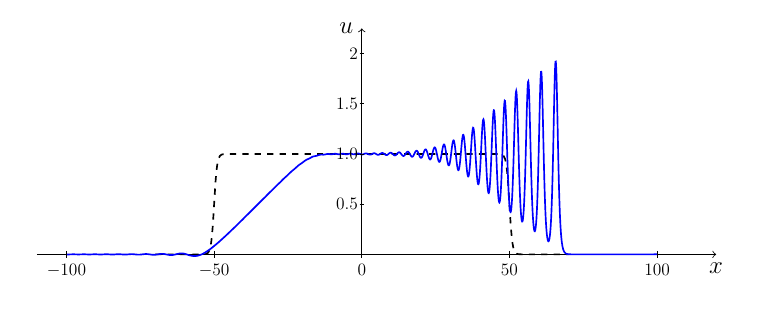}}
\caption{Evolution of the initial table-top distribution shown by a dashed line.
A rarefaction wave is formed  on the left side of the wave profile and a
dispersive shock wave on its right side.
}
\label{fig2}
\end{figure}

In considered in subsection \ref{sub4-1} problem, the initial carrier wave number was prescribed as
an `external' condition determined by a prehistory of generation and evolution of the packet
before it enters into the rarefaction wave region. In the Gurevich-Pitaevskii theory of dispersive
shock waves \cite{gp-73,eh-16,kamch-21} a shock is formed after wave breaking moment as a result
of interplay of dispersive and nonlinear properties of the physical system under consideration.
In the asymptotic Whitham approximation, the dispersive shock is formed instantly at some moment
of time from a narrow region with a steep profile around the wave
breaking point without any oscillations. Therefore it is natural to assume that in this 
approximation the initial distribution is smooth enough and the initial wave number of
the small-amplitude edge is equal to zero, $k(0)=0$. This condition determines the integration
constant $q$ in the integral of the Hamilton equations which govern the propagation of the
small-amplitude edge of the dispersive shock wave. As a result, both the wave number of a modulated
wave at this edge and its group velocity become functions of the local values of the background
variables. For example, if we consider the generalized KdV equation (\ref{eq26}) with $V_0(u)>0$,
$V(0)=0$, and wave breaking occurs at the point with $u=0$, then the integral (\ref{eq28})
transforms to
\begin{equation}\label{eq65}
  k(u)=\sqrt{\frac23V_0(u)}.
\end{equation}
The simplest application of this formula is to the step-like initial profiles. Let the dispersive shock
wave be formed at the right side of the table-top initial distribution, as is shown in Fig.~\ref{fig2}
for the KdV case with $V_0(u)=6u$. If the plateau amplitude is equal to $u_0$, then the
small-amplitude edge corresponds to $k(u_0)$ and it propagates to the left with the group velocity
\begin{equation}\label{eq66}
  v_g(k_0)=-V_0(u_0).
\end{equation}
This reproduces the results of the so-called `fitting method' developed by El \cite{el-05} which
has found a number of applications (see, e.g., 
Refs.~\cite{hoefer-14,egs-06,egkkk-07,lh-13,ep-14,ckp-16,sh-17,mfweh-20}).

\begin{figure}[th]
\centerline{\includegraphics[ width=8cm,clip]{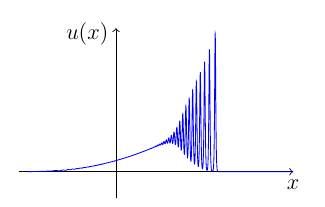}}
\caption{Evolution of a pulse after wave breaking moment. The left edge of the dispersive shock 
propagates to the left along the smooth part of the pulse and solitons are formed at the right
edge of the shock.
}
\label{fig3}
\end{figure}

\begin{figure}[th]
\centerline{\includegraphics[ width=8cm,clip]{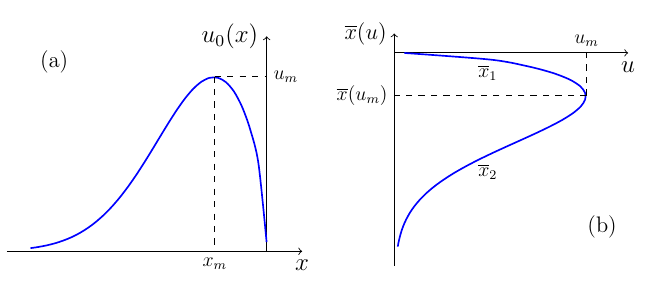}}
\caption{Initial distribution of the pulse (a) and two branches of its inverse function (b).
}
\label{fig4}
\end{figure}

Eq.~(\ref{eq55}) allows one to find the small-amplitude edge's path for an arbitrary initial profile
provided the initial condition $k(0)=0$ is known. Let the dispersive shock wave shown in Fig.~\ref{fig3}
be formed after wave breaking at $u=0$ of a pulse with the initial distribution $u=u_0(x)$ shown in
Fig.~\ref{fig4}(a). The inverse function $x=\ox(u)$ has two branches shown in Fig.~\ref{fig4}(b).
After the wave breaking moment, the small-amplitude edge starts its motion along the $\ox_1$-branch
and Eq.~(\ref{eq55}) takes the form
\begin{equation}\label{eq67}
  2V_0(u)\frac{dt}{du}+\frac{dV_0}{du}\,t=-\frac{d\ox_1}{du}.
\end{equation}
It should be solved with the initial condition $t(0)=0$ and the solution reads
\begin{equation}\label{eq68}
  t(u)=t_1(u)=-\frac1{2\sqrt{V_0(u)}}\int_0^u\frac{\ox'_1(u)}{\sqrt{V_0(u)}}\,du.
\end{equation}
Its substitution into Eq.~(\ref{eq13}) yields the path
\begin{equation}\label{eq69}
  x_1^L(u)=-\frac{\sqrt{V_0(u)}}2\int_0^u\frac{\ox'_1(u)}{\sqrt{V_0(u)}}\,du+\ox_1(u).
\end{equation}
These formulas are correct until the packet reaches the point with maximal value of $u=u_m$ at the 
moment $t_m=t_1(u_m)$. After that the packet propagates along the second branch according to
Eq.~(\ref{eq67}) with $\ox_1$ replaced by $\ox_2$ and this equation should be solved with the initial
condition $t(u_m)=t_m$. After simple calculations we obtain
\begin{equation}\label{eq70}
\begin{split}
  t(u)=&-\frac1{2\sqrt{V_0(u)}}\int_0^{u_m}\frac{\ox'_1(u)}{\sqrt{V_0(u)}}\,du\\
  &-\frac1{2\sqrt{V_0(u)}}\int_{u_m}^u\frac{\ox'_2(u)}{\sqrt{V_0(u)}}\,du,\\
  x_2^L(u)=&-\frac{\sqrt{V_0(u)}}2\int_0^{u_m}\frac{\ox'_1(u)}{\sqrt{V_0(u)}}\,du\\
  &-\frac{\sqrt{V_0(u)}}2\int_{u_m}^u\frac{\ox'_2(u)}{\sqrt{V_0(u)}}\,du+\ox_2(u),
  \end{split}
\end{equation}
These formulas determine the path $x=x^L(t)$ of the small-amplitude edge in a parametric form.
Similar formulas can be derived for other nonlinear wave equations (see, e.g., 
\cite{kamch-19,ik-19,ik-20}).

\section{Propagation of solitons}

\subsection{General theory}

The notion of the group velocity (\ref{eq10}) and existence of an integral (\ref{eq46}),
(\ref{eq48}), or (\ref{eq53}) of the Hamilton equations allowed us to develop a quite
general approach to problems of short-wavelength packets propagation. It is remarkable
that a similar approach can be developed for propagation of narrow solitons whose width
is much smaller than the characteristic length $l$ at which the background variables change,
so the position of a soliton corresponds to the coordinate $x(t)$ of its center. This approach
is based on a simple remark first made by Stokes \cite{stokes} that the exponentially 
small soliton's tails
\begin{equation}\label{eq71}
  \rho-\rho_0,u-u_0\propto\exp[-\kappa|x-Vt|]
\end{equation}
obey the same linearized equations as the harmonic linear wave (\ref{eq2}); here $V$ is the
soliton's velocity, $\kappa$ is its inverse half-width, and we assume that $\kappa^{-1}\ll l$,
so that locally the background can be considered as uniform. Then comparison of Eqs.~(\ref{eq2})
and (\ref{eq71}) gives the expression for the soliton velocity
\begin{equation}\label{eq72}
  V=\frac{\om(i\kappa,r_+,r_-)}{i\kappa},
\end{equation}
where it is assumed that $r_+,r_-$ change little across the soliton at distances about
$\kappa^{-1}\ll l$. Apparently, this formula was rediscovered several times after Stokes
(see, e.g., Refs.~\cite{schlomann,ai-77,bis-78}) and it admits a generalization on breather
solutions \cite{sazonov}. Eq.~(\ref{eq72}) replaces the formula (\ref{eq10}) in our approach 
to the theory of soliton's propagation.

Now, we assume that the condition of asymptotic integrability can also be continued to 
complex values of $k=i\kappa$. Actually, this is just an analytical continuation of
the integrals, so, depending on the dispersion relation for linear waves, we obtain 
\begin{equation}\label{eq73}
  \kappa^2=-4\sigma(q-r_+)(q-r_-),
\end{equation}
\begin{equation}\label{eq74}
  \kappa^2=-\frac{4\sigma}{q}(q-r_+)(q-r_-).
\end{equation}
Of course, this does not mean that the mapping $k\to i\kappa$ transforms the Hamilton equations 
for the packet into the Hamilton equations for the soliton considered as a point-like particle.
Instead, we have to substitute the relevant integral into Eq.~(\ref{eq72}) and obtain the
equation
\begin{equation}\label{eq76}
  \frac{dx}{dt}=\frac{\om[i\kappa(r_+(x,t),r_-(x,t)),r_+(x,t),r_-(x,t)]}{i\kappa(r_+(x,t),r_-(x,t))}
\end{equation}
for soliton's motion along the background wave described by the solution $r_+=r_+(x,t),r_-=r_-(x,t)$
of the hydrodynamic equations (\ref{eq5}).

\begin{figure}[th]
\centerline{\includegraphics[ width=8cm,clip]{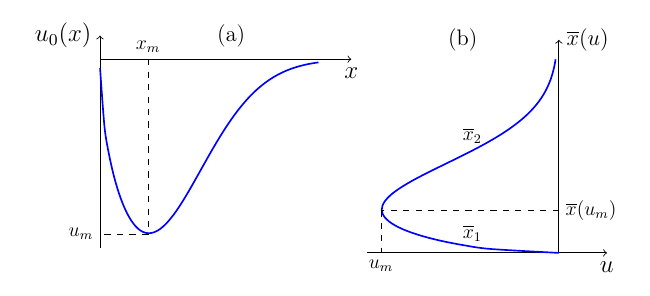}}
\caption{Initial distribution of the `negative' pulse (a) and two branches of its inverse function (b).
}
\label{fig5}
\end{figure}

The theory simplifies if the background wave is a simple wave described by a single variable
$u=u(x,t)$, so that $\kappa=\kappa(u)$ and 
\begin{equation}\label{eq77}
  V=\frac{\om(i\kappa(u),u)}{i\kappa(u)}.
\end{equation}
Differentiation of Eq.~(\ref{eq13}) with respect to $u$ yields the equation
\begin{equation}\label{eq78}
  \left[\frac{\om(i\kappa(u),u)}{i\kappa(u)}-V_0(u)\right]\frac{dt}{du}-V_0'(u)t=\ox'(u).
\end{equation}
This equation replaces Eq.~(\ref{eq55}), so its solution together with Eq.~(\ref{eq13}) gives us
the soliton's path $t=t(u),x=x(u)$ in a parametric form. For example, if we consider a `negative'
initial distribution for the generalized KdV equation (see Fig.~\ref{fig5}(a)), then the soliton
edge of the dispersive shock wave propagates along the smooth part of the evolving profile,
Eq.~(\ref{eq65}) transforms to
\begin{equation}\label{eq79}
  \kappa(u)=\sqrt{-\frac23V_0(u)}
\end{equation}
and Eq.~(\ref{eq78}) takes the form
\begin{equation}\label{eq80}
  \frac23V_0(u)\frac{dt}{du}+\frac{dV_0}{du}t=-\frac{d\ox_{1,2}}{du}.
\end{equation}
At first, the soliton edge propagates along the $\ox_1$-branch of the inverse function of the initial
distribution (see Fig.~\ref{fig5}(b)) and its path is given in a parametric form by the formulas
\begin{equation}\label{eq81}
\begin{split}
   t_1(u)=&\frac3{2(-V_0(u))^{3/2}}\int_0^u {\sqrt{-V_0(u)}}\,{\ox'_1(u)}\,du,\\
   x_1^R(u)=&\frac{-3}{2\sqrt{-V_0(u)}}\int_0^u {\sqrt{-V_0(u)}}\,{\ox'_1(u)}\,du
   +\ox_1(u).
  \end{split}
\end{equation}
After the moment $t_m=t_1(u_m)$ it moves along the $\ox_2$-branch according to the formulas
\begin{equation}\label{eq82}
\begin{split}
  &t(u)=\frac3{2(-V_0(u))^{3/2}}\int_0^{u_m} {\sqrt{-V_0(u)}}\,{\ox'_1(u)}\,du\\
  &+\frac3{2(-V_0(u))^{3/2}}\int_{u_m}^u{\sqrt{-V_0(u)}}\,{\ox'_2(u)}\,du,\\
  &x_R(u)=-\frac3{2\sqrt{-V_0(u)}}\int_0^{u_m}{\sqrt{-V_0(u)}}\,{\ox'_1(u)}\,du\\
  &-\frac3{2\sqrt{-V_0(u)}}\int_{u_m}^u{\sqrt{-V_0(u)}}\,{\ox'_2(u)}\,du+\ox_2(u).
  \end{split}
\end{equation}
At asymptotically large time $t\to\infty$ we have $u\to0$, $\ox_2(u)\to0$, so
\begin{equation}\nonumber
\begin{split}
  &t(u)\approx \frac3{2(-V_0(u))^{3/2}}\int_{-\infty}^{\infty} {\sqrt{-V_0(u_0(x))}}dx,\\
  &x_R(u)\approx -\frac3{2(-V_0(u))^{1/2}}\int_{-\infty}^{\infty} {\sqrt{-V_0(u_0(x))}}dx,
  \end{split}
\end{equation}
and we obtain the asymptotic formula
\begin{equation}\label{eq83}
\begin{split}
  &x_R\approx-\left(\frac{3A}2\right)^{2/3}t^{1/3},\\
  &A=\int_0^{\infty}\sqrt{-V(u_0(x))}\,dx
  \end{split}
\end{equation}
for the soliton edge path.

Similar results for trajectories of the soliton edge of a dispersive shock wave 
are considered in Refs.~\cite{kamch-19,ik-19,ik-20}.

\subsection{Hamiltonian theory of soliton motion for the generalized KdV equation}

As was mentioned above, the Stokes replacement $k\to i\kappa$ does not map the Hamilton
equations for the packet motion into the Hamilton equations for the soliton motion.
Nevertheless, the Hamilton theory of completely integrable equations 
\cite{zf-71,gardner-71,dickey,ft-07} suggests that the asymptotically integrable equations 
also posses the property of applicability of Hamiltonian mechanics to narrow solitons 
considered as point-like particles. Indeed, Eq.~(\ref{eq76}) can be treated as one of the
Hamilton equations of such dynamics and the knowledge of the integrals (\ref{eq73}) or
(\ref{eq74}) allows one to restore the Hamiltonian and the canonical momentum for soliton
dynamics.

To illustrate this approach by the simplest example, let us consider the generalized KdV
equation (\ref{eq56}) with the dispersion relation (\ref{eq57}), so that Eq.~(\ref{eq77})
transforms to
\begin{equation}\label{eq84}
  \frac{dx}{dt}=\frac{\om(i\kappa)}{i\kappa}=V_0(u)+\kappa^2.
\end{equation}
The integral (\ref{eq28}) for the packet motion is cast to the integral
\begin{equation}\label{eq85}
  \kappa^2=\frac23\left(q-V_0(u)\right)
\end{equation}
for the soliton motion. Differentiation of Eq.~(\ref{eq85}) along the soliton's path gives
\begin{equation}\label{eq86}
\frac{d\kappa^2}{dt}=-\frac23V_0'(u)\left(u_t+\frac{dx}{dt}u_x\right)=-\frac23V_0'(u)u_x\kappa^2,
\end{equation}
where we used Eq.~(\ref{eq84}) and eliminated $u_t$ with help of Eq.~(\ref{eq11}). We suppose
that $\kappa^2$ is a function $\kappa^2=f(p)$ of the soliton canonical momentum $p$ and
interpret Eq.~(\ref{eq84}) as the Hamilton equation,
$$
\frac{dx}{dt}=V_0(u)+f(p)=\frac{\prt H}{\prt p},
$$
so its integration yields
$$
H=V_0(u)p+\int f(p)dp.
$$
To find the function $f(p)$, we use the second Hamilton equation 
$\frac{dp}{dt}=-\frac{\prt H}{\prt x}$ or
$$
\frac{dp}{df}\frac{d\kappa^2}{dt}=-V_0'(u)u_xp,
$$
so that substitution of Eq.~(\ref{eq86}) gives $\frac{dp}{p}=\frac32\frac{df}{f}$ and $f=p^{2/3}$,
where the integration constant can be chosen equal to unity since the Hamilton equations are
invariant with respect to multiplication of $H$ and $p$ by the same constant factor. Thus, we
get the Hamiltonian for the soliton motion
\begin{equation}\label{eq87}
  H=V_0(u)p+\frac35p^{5/3}
\end{equation}
and the corresponding Hamilton equations
\begin{equation}\label{eq88}
  \frac{dx}{dt}=V_0(u)+p^{2/3},\quad \frac{dp}{dt}=-V_0'(u)u_xp.
\end{equation}
They were first derived in Ref.~\cite{mt-79} by a much more complicated method and the presented
here derivation follows mainly the approach of Ref.~\cite{ks-23}. It is worth noticing that in
case of zero background $u=0$ the Hamiltonian (\ref{eq87}) only by notation differs from the
Hamiltonian obtained in Ref.~\cite{zf-71} for a single soliton. From practical point of view,
it is important to notice that substitution of Eq.~(\ref{eq85}) into Eq.~(\ref{eq84}) gives the
equation for the soliton motion in the form
\begin{equation}\label{eq89}
  \frac{dx}{dt}=\frac13\left(V_0(u(x,t))+2q\right),
\end{equation}
where the value of $q$ is to be determined from the initial conditions. Integration of
Eq.~(\ref{eq89}) gives directly the path $x=x(t)$ of the soliton \cite{ks-23}, but the
Hamiltonian approach has its own advantages in some more complicated situations.

\subsection{Hamiltonian theory of soliton motion in general asymptotically integrable case}

In the asymptotically integrable case with two wave variables we have the formulas (\ref{eq44})
and (\ref{eq47}) for dispersion relations, the corresponding integrals (\ref{eq46}) and
(\ref{eq48}) for the packet's motion, and their counterparts (\ref{eq73}) and (\ref{eq74})
for the soliton motion. We assume for definiteness that $r_+>r_-$; then substitution of
Eq. (\ref{eq73}) or (\ref{eq74}) into the Stokes formula (\ref{eq72}) yields the equation
\begin{equation}\label{eq90}
  V=\frac{dx}{dt}=\frac12(r_+-r_-)+q
\end{equation}
for the soliton motion. It allows one to find the soliton's path $x=x(t)$ for a known
solution $r_+=r_+(x,t),r_-=r_-(x,t)$ of the hydrodynamic equations (cf. Eq.~(\ref{eq89})
for the generalized KdV equation case). Here we will use it for derivation of the
soliton's Hamiltonian and the canonical momentum assuming that a narrow soliton moves along a
known smooth background wave described by the variables $r_+(x,t),r_-(x,t)$ \cite{kamch-25}.

To be definite, we will consider the case when the integral is given by Eq.~(\ref{eq73})
with $\sigma=+1$. Then it is convenient to rewrite it in the form
\begin{equation}\label{eq91}
  \left[\frac{2q-(r_++r_-)}{r_+-r_-}\right]^2+\left(\frac{\kappa}{r_+-r_-}\right)^2=1
\end{equation}
and to introduce such a new variable $\phi$  that
\begin{equation}\label{eq92}
  \begin{split}
  & q=\frac12(r_++r_-)+\frac12(r_+-r_-)\cos\phi,\\
  & \kappa=(r_+-r_-)\sin\phi.
  \end{split}
\end{equation}
We interpret Eq.~(\ref{eq90}) as the Hamilton equation
\begin{equation}\label{eq93}
  \frac{dx}{dt}=r_++r_-+\frac12(r_+-r_-)\cos\phi=\frac{\prt H}{\prt p},
\end{equation}
and look for the canonical momentum in the form
\begin{equation}\label{eq94a}
  p=(r_+-r_-)^2f(\phi).
\end{equation}
Integration of Eq.~(\ref{eq93}) yields
\begin{equation}\label{eq95a}
  H=(r_++r_-)p+\frac12(r_+-r_-)^3\int\cos\phi f'(\phi)d\phi,
\end{equation}
and the function $f(\phi)$ can be found from the second Hamilton equation 
$\frac{dp}{dt}=-\frac{\prt H}{\prt x}$. After simple calculations (see details in 
Ref.~\cite{kamch-25}) we obtain $f(\phi)=C(\phi-\sin\phi\cos\phi)$ and we choose the
integration constant equal to $C=1/2$ for convenience of comparison with some previously
known results. As a result, we arrive at the expressions for the canonical momentum and
the Hamiltonian
\begin{equation}\label{eq94}
  p=\frac12(r_+-r_-)^2(\phi-\sin\phi\cos\phi),
\end{equation}
\begin{equation}\label{eq95}
  H=(r_++r_-)p+\frac16(r_+-r_-)^3\sin^3\phi,
\end{equation}
where $\phi$ is related to the soliton's velocity $V=\dot{x}$ by the formula
\begin{equation}\label{eq96}
  \phi=\arccos\frac{2(\dot{x}-r_+-r_-)}{r_+-r_-}.
\end{equation}
The corresponding Hamilton equations determine the soliton's dynamics. If $\sigma=-1$,
then the trigonometric functions should be replaced by the hyperbolic ones, and a similar 
theory can be developed for systems with another dispersion relation (\ref{eq47}) and
the integral (\ref{eq74}) \cite{kamch-25}. In concrete applications, the Riemann
invariants should be replaced by their expressions in terms of the physical variables.

\subsection{Application to Bose-Einstein condensate}

Let us apply the above theory to solitons in Bose-Einstein condensate (BEC) with repulsive
integration between atoms. Its dynamics obeys the Gross-Pitaevskii equation (\ref{eq14})
with $f(\rho)=\rho$ (see, e.g., \cite{ps-03}). It means that in the dispersionless
approximation the Riemann invariants are given by Eqs.~(\ref{eq22}), so their
substitution into Eqs.~(\ref{eq94})--(\ref{eq96}) yields the expressions for the
canonical momentum and the Hamiltonian of a dark soliton in BEC,
\begin{equation}\label{eq97}
\begin{split}
  p&=-2\dot{x}\sqrt{\rho_0-(\dot{x}-u)^2}+2\rho\arccos\frac{\dot{x}-u}{\sqrt{\rho}},\\
  H&=\frac43\left[\rho-(\dot{x}-u)^2\right]^{3/2}+up.
  \end{split}
\end{equation}
These expressions were first obtained in Ref.~\cite{shevchenko-88} by a different method.
The corresponding Hamilton equations are accompanied by the hydrodynamic equations for
the background large-scale wave,
\begin{equation}\label{eq98}
\begin{split}
\rho_t + (\rho u)_x = 0, \quad
u_t + uu_x + \rho_x  = -U_x,
\end{split}
\end{equation}
where we have taken into account the external force (trap potential $U(x)$) acting on BEC.
As was shown in Ref.~\cite{ik-22}, the Hamilton equations can be transformed to a more
convenient Newton-like equation
\begin{equation}\label{eq99}
\begin{split}
  &2\ddot{x}=\rho_x+(u+\dot{x})u_x+2u_t\\
  &+\frac{\rho_t+(\rho u)_x}{\sqrt{\rho-(\dot{x}-u)^2}}\arccos\frac{\dot{x}-u}{\sqrt{\rho}}.
  \end{split}
\end{equation}
The dynamics is proportional to the derivatives of $\rho$ and $u$ which are supposed to be 
small enough not to deform the soliton's profile. Elimination of $\rho_t$ and $u_t$ with
help of Eqs.~(\ref{eq98}) yields the Newton equation in the form
\begin{equation}\label{eq100}
  2\ddot{x}=-2U_x-\rho_x+(\dot{x}-u)u_x
\end{equation}
or
\begin{equation}\label{eq101}
  2\ddot{x}=-U_x+u_t+u_x\dot{x}=-U_x+\dot{u}.
\end{equation} 
The last equation was obtained in framework of the perturbation theory in Ref.~\cite{ba-2000}.
Naturally, if there is no external potential then Eq.~(\ref{eq101}) gives at once
the integral of motion
\begin{equation}\label{eq102}
  2\dot{x}-u=\mathrm{const},
\end{equation}
that is we reproduce Eq.~(\ref{eq90}). However, the account of the external potential leads
to important consequences. 
If the flow is stationary and the distributions $\rho=\rho(x)$, 
$u=u(x)$ do not depend on time $t$, then the equation
\begin{equation}\label{eq103}
  2\ddot{x}=-U_x+u_x\dot{x}
\end{equation}
has the integral of energy $H(\dot{x},x)=\mathrm{const}$. At last, if there is no flow ($u=0$) 
and a dark soliton propagates along the stationary Thomas-Fermi distribution 
$\rho(x)+U(x)=\mathrm{const}$, then its motion obeys the Newton equation
\begin{equation}\label{h7}
  2\ddot{x}=-U_x.
\end{equation}
In case of BEC confined in a harmonic trap $U(x)=\omega_0^2x^2/2$ such a soliton
oscillates with the frequency $\omega_0/\sqrt{2}$ as was predicted in Ref.~\cite{ba-2000}
and confirmed in the experiments \cite{becker-08,weller-08}.

For applications to other problems of soliton dynamics in BEC see Refs.~\cite{she-18,ik-22}.
A similar theory was developed for the DNLS equation \cite{ks-24} and for the so-called
`magnetic' solitons in a two-component BEC \cite{kamch-24c}. 

\section{Evolution of an intensive initial pulse}

As is known, an intensive enough initial pulse evolves at infinitely large time to a sequence
of solitons, provided the sign of the pulse coincides with the sign of a single soliton,
that is a positive pulse evolves into sequence of `bright' solitons and a negative pulse into
a sequence of `dark' solitons. This process is quite universal and it does not depend on complete
integrability of the equation under consideration. As was shown in 
Refs.~\cite{egs-08,kamch-20,kamch-21,cbk-21,kamch-23}, the number of asymptotic solitons and 
their parameters can be found from the initial profile of the pulse, if the equation is
asymptotically integrable and the exact or approximate integral of Hamilton equations
for a packet is known.

The approach of Refs.~\cite{kamch-20,kamch-21,cbk-21} was based on a simple remark made by 
Gurevich and Pitaevskii in Ref.~\cite{gp-87} that oscillations
enter into the dispersive shock region with velocity equal to difference between the group and
phase velocities of this small-amplitude edge wave packet, 
\begin{equation}\label{eq107}
  \frac{dN_{\text{DSW}}}{dt}=\frac1{2\pi}|k(v_g-V)|,
\end{equation}
where the wave number $k$, the group velocity $v_g=\prt\om/\prt k$, and the phase velocity
$V_{ph}=\om/k$ refer to the instant location of this edge at the moment $t$. This formula
can be written \cite{kamch-21} as
\begin{equation}\label{eq108}
  \frac{dN_{\text{DSW}}}{dt}=
  \frac1{2\pi}\left|k\frac{\prt\om}{\prt k}-\om\right|,
\end{equation}
and then it can be interpreted as a direct consequence of the number of waves conservation
law (\ref{eq25}) written in the reference frame moving with the group velocity of the edge,
that is the flux of waves $\om$ is changed by the Doppler shift  $kv_g$.
Then the total number of waves generated in the dispersive shock wave from a large pulse
is equal to
\begin{equation}\label{eq109}
  N=\frac1{2\pi}\int_0^{\infty}\left|k\frac{\prt\om}{\prt k}-\om\right|dt.
\end{equation}
The integration is taken along the total 
path of the small-amplitude edge from the wave breaking moment $t=0$ till this edge reaches
the opposite side of the pulse with vanishing intensity. Eventually, each crest of the
dispersive shock wave evolves to a soliton, so Eq.~(\ref{eq109}) yields the total number of
solitons produced from an initial pulse.

\subsection{Number of solitons in the generalized KdV equation theory}

In case of the generalized KdV equation (\ref{eq26}) with $V_0(u)>0$, a large initial pulse
$u=u_0(x)$ leads to formation of $N$ solitons and this number can easily be calculated
with the use of Eq.~(\ref{eq109}), since the wave number $k$ is given by Eq.~(\ref{eq65})
and the path of the small-amplitude edge by Eqs.~(\ref{eq68})-(\ref{eq70}). Their
substitution into Eq.~(\ref{eq109}) and simple transformations yield
\begin{equation}\label{eq110}
\begin{split}
  N=\frac{(2/3)^{3/2}}{2\pi}&\Bigg\{\int_0^{u_m}du\frac{V_0'(u)}2\int_u^{u_m}
  \frac{(\ox_2'-\ox_1')du_1}{\sqrt{V_0(u_1)}}\\
 & +\int_0^{u_m}\sqrt{V_0(u)}(\ox_2'-\ox_1')du\Bigg\}.
  \end{split}
\end{equation}
Here, the double integral reduces to the ordinary one by integration
by parts with account of $V_0(0)=0$, so we arrive at the expression
\begin{equation}\label{eq111}
\begin{split}
  N&=\frac1{2\pi}\int_0^{u_m}\sqrt{\frac23V_0(u)}\,(\ox_2'-\ox_1')\,du\\
  &=\frac1{2\pi}\int\sqrt{\frac23V_0(u_0(x))}\,\,dx.
  \end{split}
\end{equation}
At last, remembering Eq.~(\ref{eq65}), we write this formulas as
\begin{equation}\label{eq112}
  N=\frac1{2\pi}\int k[u_0(x)]dx.
\end{equation}
Of course, this formula is only correct for a large number of solitons $N\gg1$.

\begin{figure}
\includegraphics[width=8cm]{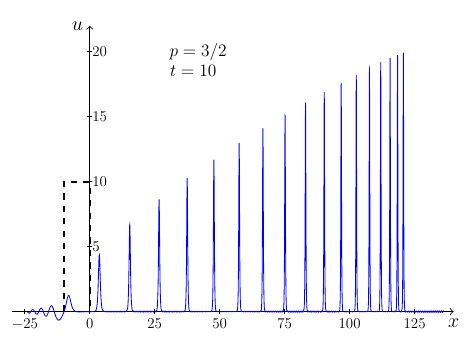}
\caption{The profile of $u(x)$ at $t=10$ obtained by numerical solution
of Eq.~(\ref{eq26}) with the table-top initial condition (\ref{eq116})
shown by a dashed line (from Ref.~\cite{kamch-20}).}
\label{fig6}
\end{figure}

We illustrate the above theory by the generalized KdV equation with
\begin{equation}\label{eq113}
  V_0(u)=6u^p,
\end{equation}
when Eq.~(\ref{eq26}) has a soliton solution ($p>0$),
\begin{equation}\label{eq114}
  u_s(x,t)=\frac{u_m}{\cosh^{2/p}\left[\tfrac12p\sqrt{V_s}(x-V_st)\right]},
\end{equation}
where
\begin{equation}\label{eq115}
  u_m=\left[\frac1{12}V_s(p+1)(p+2)\right]^{1/p}.
\end{equation}
As was shown in Ref.~\cite{kuzn-84}, these solitons are unstable for $p>4$,
so we confine ourselves to the cases with $p<4$. 

For $p=1$ and $p=2$ we return to the completely KdV and mKdV equations,
respectively, and for these equations the formula (\ref{eq112}) can be obtained by the 
IST method \cite{karpman-67,karpman-73}. Therefore we
will consider non-integrable cases $p=3/2$ and $p=1/2$ (so
called Schamel equation, Ref.~\cite{schamel-73}). The initial pulse is
taken in the table-top form
\begin{equation}\label{eq116}
  u_0(x)=\left\{
  \begin{array}{ll}
  u_0,\quad &-l\leq x\leq 0,\\
  0,\quad &x<-l\quad\text{and}\quad x>0,
  \end{array}
  \right.
\end{equation}
so that Eq.~(\ref{eq112}) reduces to the asymptotic formula $(N\gg1)$
\begin{equation}\label{eq117}
  N\approx\frac{l}{\pi}u_0^{p/2}.
\end{equation}

\begin{figure}
\includegraphics[width=8cm]{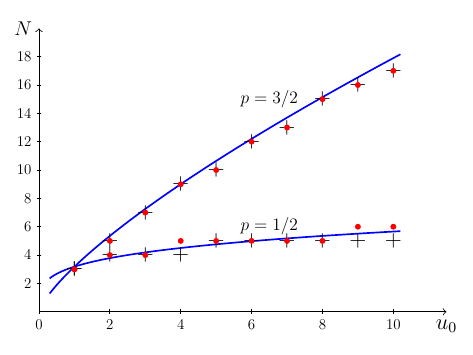}
\caption{Number of solitons produced from the table-top initial pulse
(\ref{eq116}) with $l=10$ and integer values of $u_0$ from 1 to 10.
Solid lines corresponds to the plots of the function (\ref{eq117}), dots
give its rounded to integer values, and crosses correspond to the
numerical solutions of Eq.~(\ref{eq26}) (from Ref.~\cite{kamch-20}). }
\label{fig7}
\end{figure}

In Fig.~\ref{fig6} a typical resulting profile is shown for $p=3/2$, $l=10$
and evolution time $t=10$. Solitons have positive velocities and
are located in the region $x>0$. The dependence of $N$ on $u_0$
is shown in Fig.~\ref{fig7} where the solid lines depict the plots of the function
$N(u_0)$. As one can see, the agreement of theoretical predictions with numerical 
results is very good even in this case of not completely integrable equations.

It is remarkable that the same formula (\ref{eq112}) is obtained for many other equations
\cite{cbk-21,kamch-21b}, if a simple wave pulse is considered. This observation suggests
that this formula has some general origin.

\subsection{Poincar\'{e}-Cartan integral invariant}\label{P-C}

Equation (\ref{eq109}) can be written in the form
\begin{equation}\label{eq118}
  N=\frac{S}{2\pi},
\end{equation}
where $S=\int(k\delta x-\om \delta t)$ is the action of a classical particle associated with the 
small-amplitude edge packet. In this formula, the action is to be calculated along
the path of the small-amplitude edge and the result (\ref{eq112}) of this calculation 
is again the action integral but calculated now along the path associated with the 
initial condition at $t=0$. If we introduce the inverse path  with $k$ replaced by $-k$, 
then the combined direct and inverse paths make a closed contour $C$ in the phase space $(x,k)$. 
As a result, Eq.~(\ref{eq118}) can be written in
the form of the Poincar\'e-Cartan integral invariant (see, e.g. Ref.~\cite{gant-66})
\begin{equation}\label{eq119}
  N=\frac1{4\pi}\oint_C(k\delta x-\om\delta t),
\end{equation}
where the additional factor $1/2$ is introduced for taking into
account that in the case of a closed contour the edge goes along its path twice.
Then the above calculation suggests that this integral is preserved by the simple-wave 
hydrodynamic flow (\ref{eq11}). As was shown in Ref.~\cite{kamch-23}, this suggestion
is true, if $k$ is given by the solution (\ref{eq24}) of Eq.~(\ref{eq23}). Moreover,
this statement can be extended to the general hydrodynamic flows obeying Eqs.~(\ref{eq5})
provided the system under consideration is asymptotically integrable and $k$ is a function
of the background wave variables given by the solution
\begin{equation}\label{eq120}
  k=k(r_+,r_-,q)
\end{equation}
of Eqs.~(\ref{eq42}) (see, e.g., (\ref{eq46}) or (\ref{eq48})). Consequently, we can transform
the contour $C$ to the initial state $t=0$ and then Eq.~(\ref{eq117}) gives the expression
\begin{equation}\label{eq121}
  N=\frac{1}{2\pi}\int_{-\infty}^{\infty}k(r_+^0(x),r_-^0(x),q)dx,
\end{equation}
where $r_{\pm}^0(x)$ are the initial distributions of the dispersionless Riemann invariants
and $q$ is chosen according to the initial condition $k=0$ at the wave breaking moment.

\subsection{Generalized Bohr-Sommerfeld quantization rule}

In fact, integration in Eq.~(\ref{eq121}) is taken over an interval between two `turning
points' where the integrand is real. It coincides with asymptotic
formulas obtained from the quasiclassical Bohr-Sommerfeld quantization rule for KdV
equation in Refs.~\cite{karpman-67,karpman-73}, for NLS equation in Refs.~\cite{JLML-99,kku-02},
and for the Kaup-Boussinesq equation in Ref.~\cite{kku-03}. Therefore, Eq.~(\ref{eq121}) can be
called a generalized Bohr-Sommerfeld quantization rule applicable to the asymptotically
integrable equations \cite{kamch-23}. We define the turning points $x_{1,2}(q)$ by the equation 
\begin{equation}\label{eq122}
  k[r_+(x_{1,2}(q)),r_-(x_{1,2}(q)),q]=0
\end{equation}
and write this rule in the form 
\begin{equation}\label{eq123}
\begin{split}
  &\frac{1}{2\pi}\int_{x_1(q_n)}^{x_2(q_n)}k[r_+^0(x),r_-^0(x),q_n)dx= n+\frac12,\\
  &n=1,2,\ldots,N,
  \end{split}
\end{equation}
where the integer number $n$ is replaced by $n+\frac12$ in accordance with the well-known
correction to the quasiclassical approximation (see, e.g., \cite{LL-3,arnold}).
Actually, this is the equation for finding such `eigenvalues' $q_n$ that the expression in
the left-hand side takes half-integer values.

The biggest possible value $n=N$ corresponds to the total number of solitons produced from an intensive
initial pulse or, in other words, to the $N$th soliton in the asymptotic soliton train. In a
similar way, we can find the values $q_{N-1},q_{N-2},\ldots,q_1$, and relate each
eigenvalue $q_n$ obtained from the Bohr-Sommerfeld quantization rule (\ref{eq123}) to the
$n$th soliton in the asymptotic train. For example, if we take the NLS equation with $r_{\pm}$
given by Eqs.~(\ref{eq22}) and $k^2=(q-u)^2-4\rho]$ (see Eq.~(\ref{eq51})), then we get at once that the
Bohr-Sommerfeld rule reads
\begin{equation}\label{eq124}
\begin{split}
  &\frac{1}{\pi}\int_{x_1(q_n)}^{x_2(q_n)}\sqrt{\frac14\left(q_n-u_0(x)\right)^2-\rho_0(x)} dx= n+\frac12,\\
  &n=1,2,\ldots,N,
  \end{split}
\end{equation}
in agreement with the results of Refs.~\cite{JLML-99,kku-02} obtained in the framework of the inverse
scattering transform method. The parameters $q_n$ are related to the asymptotic solitons velocities
$V_n$ by the Stokes formula (\ref{eq72}),
\begin{equation}\label{eq125}
  V_n=\frac{\om(i\kappa_n,r_+^{\infty},r_-^{\infty})}{i\kappa_n},
\end{equation}
where $\kappa_n$ is defined by Eq.~(\ref{eq73}) with $q_n\to q_n/2$,
\begin{equation}\label{eq126}
  \kappa_n^2=-4(q_n/2-r_+^{\infty})(q_n/2-r_-^{\infty}),
\end{equation}
and $r_{\pm}^{\infty}$ represent the values of the background wave at the asymptotically large
time $t\to\infty$; usually they are equal to the background values far enough from the initial pulse.
If the asymptotic solitons propagate along the uniform state with $\rho_0(\infty)=\rho_0$,
$u_0(\infty)=0$, we get $r_{\pm}^{\infty}=\sqrt{\rho_0}$, $\kappa_n^2=4\rho_0-q_n^2$, and
\begin{equation}\label{eq127}
  V_n=\sqrt{\rho_0-\frac{\kappa_n^2}{4}}=\frac{q_n}2.
\end{equation}

If we only know an approximate solution
\begin{equation}\label{eq128}
  k=K(r_+,r_-,q)
\end{equation}
of the asymptotic integrability equations (see, e.g., (\ref{eq53})), then we cannot continue
it directly to the region of complex $k=i\kappa$. However, we know that the asymptotic
solitons produced from a large initial pulse belong to the right- or left-propagating simple
wave, and in this case we can always solve Eq.~(\ref{eq23}) and find the function (see Eq.~(\ref{eq24}))
\begin{equation}\label{eq129}
  k=\overline{K}(r_+,\overline{q}),
\end{equation}
where for definiteness we assumed that the simple wave corresponds to $r_-=r_-^0=\mathrm{const}$
and changed the notation for the integration constant $q\to\overline{q}$ to distinguish it from
the constant $q$ appeared in the general case (\ref{eq128}). The solutions (\ref{eq128}) with 
$r_-=r_-^0$  overlaps with (\ref{eq129}) in the region where Eq.~(\ref{eq128}) is correct with good 
enough accuracy.
This matching condition allows one to find the relationship between $\overline{q}$ and $q$,
$\overline{q}=\overline{q}(q)$, that is
\begin{equation}\label{eq130}
  \overline{q}_n=\overline{q}(q_n).
\end{equation}
Then $q_n$ found from the Bohr-Sommerfeld quantization rule determines the value of $\overline{q}_n$
which allows us to find the asymptotic velocities of solitons,
\begin{equation}\label{eq131}
  V_n=\frac{\om[i\overline{K}(\overline{q}_n,r_+^{\infty}),\overline{q}_n]}
  {i\overline{K}(\overline{q}_n,r_+^{\infty})}.
\end{equation}

Let us illustrate this theory by a concrete example \cite{kamch-23}. In the case of the generalized
NLS equation (\ref{eq14}) with $f(\rho)$ given by Eq.~(\ref{eq36}), which is not completely integrable
for $p\neq 1$, we get an approximate integral (see Eq.~(\ref{eq53}))
\begin{equation}\label{eq133}
  k^2=(q-u)^2-2\left(1-\frac{1}{p}\right)c^2,
\end{equation}
so the Bohr-Sommerfeld quantization rule reads
\begin{equation}\label{eq134}
\begin{split}
  &\int_{x_1(q_n)}^{x_2(q_n)}\sqrt{(q_n-u_0(x))^2-2\left(1-\frac{1}{p}\right)c_0^2(x)}\,dx\\
  &=2\pi\left(n+\frac12\right),\qquad   n=1,2,\ldots,N.
   \end{split}
\end{equation}
The dispersionless Riemann invariants are equal to 
\begin{equation}\label{eq135}
  r_{\pm}=\frac{u}{2}\pm\frac{c}{p},
\end{equation}
so for a simple wave with the constant Riemann invariant
$$
r_-=\frac{u}{2}-\frac{c}{p}=-\frac{c_R}{p}=\mathrm{const}
$$
Eq.~(\ref{eq133}) in the limit $k\sim q\gg c$ gives the expansion
\begin{equation}\label{eq136}
\begin{split}
  k^2=&\left(q+\frac{2c_R}{p}\right)^2-\frac{4}{p}\left(q+\frac{2c_R}{p}\right)c\\
  &+\left(\frac{4}{p^2}-\frac{2}{p}-2\right)c^2,\qquad c\ll q.
  \end{split}
\end{equation}
On the other hand, the exact simple wave solution case (\ref{eq37}) in the same limit $\al\gg1$ yields
\begin{equation}\label{eq137}
  \frac{c}{\oq}=\beta(p)\left(\frac{1}{\al}-\frac{1}{p}\cdot\frac{1}{\al^2}+
  \frac{p^2-p+4}{4p^2}\cdot\frac{1}{\al^3}+\ldots\right),
\end{equation}
where
\begin{equation}\label{eq138}
  \beta(p)=2^{\frac{2-p}{3p-2}}\left(1+\frac{2}{p}\right)^{\frac{2(p-1)}{3p-2}}.
\end{equation}
Inversion of this series and substitution of the result into Eq.~(\ref{eq34}) gives
\begin{equation}\label{eq139}
\begin{split}
  k^2\approx &(2\oq\beta)^2-\frac{4}{p}(2\oq\beta)c\\
  &+\left(\frac{4}{p^2}-\frac{2}{p}-2\right)c^2,\qquad c\ll\oq.
  \end{split}
\end{equation}
Comparison of Eqs.~(\ref{eq136}) and (\ref{eq139}) gives the relationship
\begin{equation}\label{eq140}
  \oq=\frac{1}{2\beta}\left(q+\frac{2c_R}{p}\right).
\end{equation}
The integral for the simple wave background wave can be continued to the soliton region 
$k\to i\kappa$, so we get
\begin{equation}\label{eq141}
\begin{split}
  V_n=c_R\al\left(c_R,\frac{1}{2\beta}\left(q_n+\frac{2c_R}{p}\right)\right),
  \end{split}
\end{equation}
that is the asymptotic solitons velocities are expressed as functions of the eigenvalues
$q_n$ found from the generalized Bohr-Sommerfeld quantization rule (\ref{eq134}).

\begin{figure}[t]
\begin{center}
	\includegraphics[width = 8cm,height = 6cm]{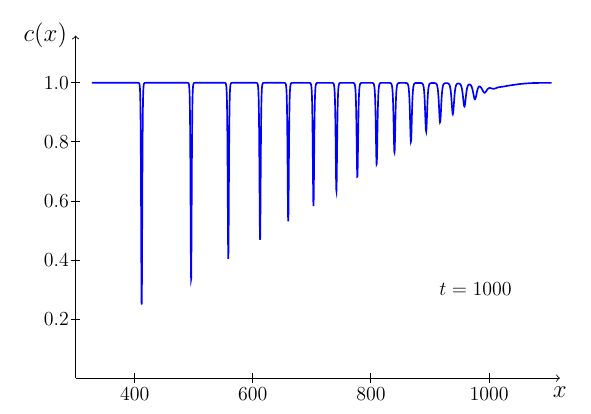}
\caption{Distribution of $c(x)$ in the right-propagating soliton train evolved from
the pulse with the initial distributions (\ref{eq144}) at $t=1000$ (from Ref.~\cite{kamch-23}).
 }
\label{fig8}
\end{center}
\end{figure}

We illustrate the theory by a concrete example with $p=2$, when $\beta=\sqrt{2}$ and (see Eq.~(\ref{eq40}))
\begin{equation}\label{eq142}
  \al(c,\oq)=\frac12\left(\sqrt{1+8\left(\frac{\oq}{c}\right)^2}-1\right).
\end{equation}
In this case Eq.~(\ref{eq141}) reduces to
\begin{equation}\label{eq143}
  V_n=\frac12\left(\sqrt{c_R^2+(c_R+q_n)^2}-c_R\right).
\end{equation}
If we take the initial distributions
\begin{equation}\label{eq144}
  c_0(x)=1-\frac{0.9}{\cosh^2(x/20)},\quad u_0(x)=0,
\end{equation}
then this initial pulse evolves into two symmetrical right- and left-propagating dark
soliton trains. A typical distribution of $c(x)$ at $t=1000$ is shown in Fig.~{\ref{fig8}
for the right-propagating solitons whose positions are given by an approximate formula
\begin{equation}\label{eq145}
  x_n(t)\approx V_nt=\frac12(\sqrt{1+(1+q_n)^2}-1)\,t.
\end{equation}
Thus, we can find $q_n^{\text{num}}$ from numerically calculated positions taken from
Fig.~\ref{fig8} and compare them with the eigenvalues $q_n^{\text{B.-S.}}$ obtained from
the Bohr-Sommerfeld quantization rule with the distributions (\ref{eq144}). The results are shown
in Fig.~\ref{fig9} and they demonstrate quite a good agreement. Thus, the generalized
Bohr-Sommerfeld rule works perfectly well even for not completely integrable equations.

\begin{figure}[t]
\begin{center}
	\includegraphics[width = 8cm,height = 6cm]{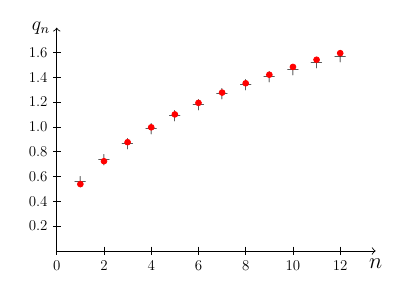}
\caption{Parameters $q_n^{\text{B.-S.}}$ obtained from the Bohr-Sommerfeld rule (crosses)
and $q_n^{\text{num}}$ obtained from the numerical solution (red dots) 
(from Ref.~\cite{kamch-23}).
 }
\label{fig9}
\end{center}
\end{figure}

\section{Connection with complete integrability}

As was indicated in Section~\ref{P-C}, the value of the Poincar\'{e}-Cartan integral invariant
is preserved by the hydrodynamic background flow and, moreover, the eigenvalues calculated
from the Bohr-Sommerfeld quantization rule can be continued to the asymptotic stage of evolution
when they yield the soliton velocities. This is a strong indication that these eigenvalues
can be interpreted as asymptotic approximation to the exact spectrum of some linear operator
and this spectrum is preserved by the nonlinear wave evolution according to the equation
under consideration. In this sense, the asymptotic integrability leads to the properties of
nonlinear wave equations which are usually related to their complete integrability. It is
natural to ask if it is possible to restore the Lax pair from the additional integrals obtained
as solutions of equations of asymptotic integrability. In fact, these integrals (\ref{eq46})
or (\ref{eq48}) can be interpreted as expressions for the `momenta' of some mechanical
systems, and then the Bohr-Sommerfeld rule (\ref{eq123}) is nothing but the quantization
rule of the `old quantum theory' applied to these mechanical systems (see, e.g., 
\cite{born,tomonaga}). Consequently, we have to
`quantize' these mechanical systems in order to go beyond the asymptotic short wavelength
approximation and to arrive at the `full quantum theory' described by the Lax pairs 
corresponding to the completely integrable
equations. In this section, we will show that indeed we can arrive in this way at the Lax pairs 
for some classes of completely integrable equations. To this end, we will use the well-known 
classical quantization methods of Schr\"{o}dinger~\cite{schr-26,schr-26b} for scalar particles 
and of Dirac~\cite{Dirac} for spinor particles. 

\subsection{KdV equation}

Let the background wave evolution be described by the Hopf equation (\ref{eq11}) and the linear
waves obey the dispersion relation (\ref{eq27}), so the integral is given by Eq.~(\ref{eq28}).
Following the Schr\"{o}dinger quantization rule, we make a replacement 
\begin{equation}\label{eq146a}
  k\to-2i\frac{\prt}{\prt x},
\end{equation}
where the factor 2 is introduced for convenience of comparison with the known results, and
convert Eq.~(\ref{eq28}) into the differential equation for the function $\phi(x)$,
\begin{equation}\label{eq146}
  \phi_{xx}=\mathcal{A}\phi,
\end{equation}
where 
\begin{equation}\label{eq146b}
  \mathcal{A}=-\frac16(V_0(u)-q),
\end{equation}
and $q$ plays the role of the eigenvalue.

The wave variable $u=u(x,t)$ depends also on time $t$, so the eigenfunction $\phi$ must
depend on $t$, too. Assuming that this dependence is governed by a linear equation
$\phi_t=\mathcal{C}\phi+\mathcal{B}\phi_x$ (higher order derivatives can be excluded with the
help of Eq.~(\ref{eq146})), we find from the compatibility condition $(\phi_{xx})_t=(\phi_t)_{xx}$
that $\mathcal{C}=-\frac12\mathcal{B}_x$, that is
\begin{equation}\label{eq147}
  \phi_t=-\frac12  \mathcal{B}_x\phi+ \mathcal{B}\phi_x,
\end{equation}
and then
\begin{equation}\label{eq148}
\mathcal{A}_t-2\mathcal{B}_x\mathcal{A}-\mathcal{B}\mathcal{A}_x+
\frac12\mathcal{B}_{xxx}=0.
\end{equation}
If $\mathcal{B}$ is known, then this equation converts into the evolution equation for the
wave variable.

To find the function $\mathcal{B}$, we notice that the second order equation (\ref{eq146})
has two basis solutions $\phi_+,\phi_-$ and their product $g=\phi_+\phi_-$ satisfies
the equation 
\begin{equation}\label{eq149}
 \left(\frac1{g}\right)_t-\left(\frac{\mathcal{B}}{g}\right)_x=0.
\end{equation}
On the other hand, the basis solutions of Eq.~(\ref{eq146}) can be expressed in terms of $g$
by the formula
\begin{equation}\label{eq149b}
  \phi_{\pm}=\sqrt{g}\exp\left(\pm i\int^x\frac{dx'}{g}\right),
\end{equation}
that is $1/g$ has the meaning of the wave number of a slightly modulated wave.
Consequently, Eq.~(\ref{eq149}) must coincide with the number of waves
conservation law (\ref{eq25}), so  we have
\begin{equation}\label{eq150}
  k=\frac1g,\quad \mathcal{B}=-\frac{\om}{k}.
\end{equation}
In our case with $\om=V_0k-k^3$ and $k^2=\frac23(V_0-q)$ we assume that $\mathcal{B}$ 
depends only on the wave variable $u$, as it takes place for the function $\mathcal{A}$ in 
(\ref{eq146b}), and get
\begin{equation}\label{eq151}
  \mathcal{B}=-\frac13(V_0(u)+2q).
\end{equation}
Substitution of Eqs.~(\ref{eq146b}) and (\ref{eq151}) into Eq.~(\ref{eq148}) yields
\begin{equation}\label{eq152}
  \frac{\prt V_0(u)}{\prt t}+V_0(u)\frac{\prt V_0(u)}{\prt x}+\frac{\prt^3 V_0(u)}{\prt x^3}=0.
\end{equation}
This is actually the KdV equation for the variable $V_0(u)$. Linearization of this equation
reproduces the dispersion relation (\ref{eq27}), as it should be, but if Eq.~(\ref{eq152})
is written for the variable $u$,
\begin{equation}\label{eq153}
  u_t+V_0'(u)u_x+u_{xxx}+\frac{3V_0^{\prime\prime}(u)}{V_0'(u)}u_xu_{xx}+
  \frac{V_0^{\prime\prime\prime}(u)}{V_0'(u)}u_x^3=0,
\end{equation}
then we get a more complicated nonlinear dispersion terms. In spite of them, Eq.~(\ref{eq153})
is just a disguised form of the KdV equation (\ref{eq152}).

Strictly speaking, the number of waves conservation law (\ref{eq25}) is asymptotic and 
Eq.~(\ref{eq150}) is only correct when both $\mathcal{A}$ and $\mathcal{B}$ do not depend on 
the $x$-derivatives of the wave variable. Otherwise the formulas (\ref{eq146b}) and (\ref{eq150}) 
only provide quasiclassical approximations to the exact expressions for $\mathcal{A}$ and
$\mathcal{B}$ (see Ref.~\cite{ks-24a}). For example, the Lax pair for the completely integrable 
modified KdV equation
\begin{equation}\label{eq154}
  u_t+6u^2u_x+u_{xxx}=0
\end{equation}
is usually written in a matrix form \cite{wadati-73,nmpz-80} and its transformation to a scalar
form \cite{kk-02} yields
\begin{equation}\label{eq155}
  \begin{split}
  & \mathcal{A}=\left(\la+\frac{u_x}{2u}\right)^2-u^2-\left(\frac{u_x}{2u}\right)_x,\\
  & \mathcal{B}=-4\la^2-2u^2+2\la\frac{u_x}{u}-\frac{u_{xx}}{u}.
  \end{split}
\end{equation}
It is easy to see that in the quasiclassical limit when the terms containing derivatives of $u$
are neglected, we obtain the formulas
\begin{equation}\label{eq156}
  \overline{\mathcal{A}}=\la^2-u^2,\quad \overline{\mathcal{B}}=-4\la^2-2u^2,
\end{equation}
which coincide with Eqs.~(\ref{eq146b}) and (\ref{eq151}), if we take $V_0(u)=6u^2$ and $q=6\la^2$.
However, in order to get the full Lax pair (\ref{eq155}), one should use some other method
of quantization.

\subsection{Lax pairs with `energy dependent potentials'}

As was indicated above in the generalized KdV equation case, the Schr\"{o}dinger quantization rule
leads to the Lax pair, which does not depend on derivatives of the wave variable but does depend on
the `spectral parameter' $q$. Here we will extend this scheme to the asymptotically integrable equations 
with two wave variables. The corresponding Lax pairs are called `pairs with energy-dependent potentials' 
\cite{jm-76,pavlov-14}. In this case, since
the additional integral is only a function of the Riemann invariants, but not of their derivatives,
the Schr\"odinger quantization rule (\ref{eq146a}) yields again the spectral problem (\ref{eq146}) 
with the Lax pairs in the form
\begin{equation}\label{k1}
  \begin{split}
  \mathcal{A}=-\frac14k^2,\qquad
  \mathcal{B}=-\frac{\om}{k}.
  \end{split}
\end{equation}
We know two different integrals (\ref{eq46}) and (\ref{eq48}) which correspond respectively to two
different dispersion relations (\ref{eq44}) and (\ref{eq47}). Consequently, we get the functions
$\mathcal{A}$ given by
\begin{equation}\label{eq158}
  \mathcal{A}=-\sigma(q-r_+)(q-r_-),
\end{equation}
or
\begin{equation}\label{eq159}
  \mathcal{A}=-\frac{\sigma}{q}(q-r_+)(q-r_-).
\end{equation}
In both cases Eq.~(\ref{k1}) gives
\begin{equation}\label{eq160}
  \mathcal{B}=-q-\frac12(r_++r_-).
\end{equation}
Then the compatibility condition (\ref{eq148}) yields the system
\begin{equation}\label{eq161}
  \begin{split}
   (r_++r_-)_t&+\frac32(r_++r_-)(r_++r_-)_x
      -(r_+r_-)_x=0,\\
   (r_+r_-)_t&+r_+r_-(r_++r_-)_x
   +\frac12(r_++r_-)(r_+r_-)_x\\ &   +\frac{\sigma}{4}(r_++r_-)_{xxx}=0
  \end{split}
\end{equation}
in case of $\mathcal{A}$ given by Eq.~(\ref{eq158}), and the system
\begin{equation}\label{eq162}
  \begin{split}
   (r_++r_-)_t&+\frac32(r_++r_-)(r_++r_-)_x\\
   &-(r_+r_-)_x-\frac{\sigma}{4}(r_++r_-)_{xxx}=0,\\
   (r_+r_-)_t&+r_+r_-(r_++r_-)_x\\
   &+\frac12(r_++r_-)(r_+r_-)_x=0,
  \end{split}
\end{equation}
in case of $\mathcal{A}$ given by Eq.~(\ref{eq159}).

These completely integrable systems get physical sense when some concrete expressions for
the Riemann invariants are substituted. For example, if we substitute (\ref{eq22})
into Eqs.~(\ref{eq161}), then we get the Kaup-Boussinesq system \cite{bouss-1877,kaup-75}
\begin{equation}\label{eq163}
  \rho_t+(\rho u)_x-\frac{\sigma}4u_{xxx}=0,\quad u_t+uu_x+\rho_x=0.
\end{equation}
On the other hand, substitution of
\begin{equation}\label{eq165}
  r_{\pm}=\frac{u}{2}\pm\sqrt{\rho+\frac{u^2}{4}}
\end{equation}
gives the Jaulent-Miodek system \cite{jm-76}
\begin{equation}\label{eq166}
\begin{split}
  & u_t+\frac32uu_x+\rho_x=0,\\ 
  & \rho_t+\rho u_x+\frac12\rho_x-\frac{\sigma}{4}u_{xxx}=0.
  \end{split}
\end{equation}
At last, if we substitute
\begin{equation}\label{eq164}
  r_{\pm}=\frac{u}{2}\pm\sqrt{\frac{u^2}{4}-\rho}
\end{equation}
into Eqs.~(\ref{eq162}), then we obtain the Zakharov-Ito system \cite{zakh-80,ito-82}
\begin{equation}\label{eq167}
  \begin{split}
   &u_t+\frac32uu_x-\rho_x-\frac{\sigma}4u_{xxx}=0,\\
   &\rho_t+\rho u_x+\frac12u\rho_x=0.
  \end{split}
\end{equation}
Other examples can be found in Ref.~\cite{kamch-25}. Obviously, the systems which belong
to either of the general forms (\ref{eq161}) or (\ref{eq162}) can be transformed one to
another by simple algebraic formulas relating the wave variables.

\subsection{Gross-Pitaevskii (NLS) equation}

Suppose we are interested in finding a dispersive generalization of Landau hydrodynamics
\cite{landau-41,landau-47} of superfluid gas described by the equations
\begin{equation}\label{eq168}
  \rho_t + (\rho u)_x = 0, \quad
u_t + uu_x + \rho_x  =0.
\end{equation}
From Bogoliubov paper \cite{bogol-47} we know the dispersion relation
\begin{equation}\label{eq169}
  \om=k\left(u+\sqrt{\rho+\frac{k^2}{4}}\right)
\end{equation}
for linear waves propagating along a uniform state. This dispersion relation coincides with
Eq.~(\ref{eq44}), if we substitute the Riemann invariants (\ref{eq22}) corresponding to
Eqs.~(\ref{eq168}), so our system is asymptotically integrable and the integral (\ref{eq46})
with $\sigma=+1$ takes the form (\ref{eq51}). The Schr\"{o}dinger quantization rule leads to 
the Kaup-Boussinesq system corresponding to dynamics of a classical fluid. However, in case
of quantum fluids, the existence of quantized vortices in liquid Helium~II 
\cite{onsager,feynman-55} suggests that their states should be described by such a complex
function $\psi$ that the gradient of its phase is proportional in the quasiclassical limit 
to the fluid velocity $u$. This means that $\rho$ and $u$ must be combined into a complex 
wave function $\psi$, and we assume that this remains true in our one-dimensional case.
Therefore, instead of the Schr\"odinger rule, we will use the quantization method of
Dirac~\cite{Dirac} according to which the integral is a consequence of a linear system
of first-order differential equations,
\begin{equation}\label{eq170}
  \begin{split}
  & \frac{\prt\tilde{\phi}_1}{\prt x}=\frac{i}{2}(q-u)\tilde{\phi}_1+i\sqrt{\rho}\tilde{\phi}_2,\\
  & \frac{\prt\tilde{\phi}_2}{\prt x}=-i\sqrt{\rho}\tilde{\phi}_1-\frac{i}{2}(q-u)\tilde{\phi}_2.
  \end{split}
\end{equation}
Then for the plane wave $$
\left(
\begin{array}{c}
\tilde{\phi}_1 \\
\tilde{\phi}_2 \\
\end{array}
\right) = \left(
                                     \begin{array}{c}
                                       a_1 \\
                                       a_2 \\
                                     \end{array}
                                   \right)e^{ikx/2}
$$
we obtain the relationship (\ref{eq51}) which becomes a quasiclassical consequence of this system.
Now we make the replacements
\begin{equation}\label{eq171}
\begin{split}
  & \tilde{\phi}_1=\phi_1\exp\left(-\frac{i}{2}\int^xudx'\right),\\
  & \tilde{\phi}_2=\phi_2\exp\left(\frac{i}{2}\int^xudx'\right),
  \end{split}
\end{equation}
so Eqs.~(\ref{eq170}) take the form
\begin{equation}\label{eq172}
  \begin{split}
  & \frac{\prt{\phi}_1}{\prt x}=\frac{i}{2}q{\phi}_1+i\sqrt{\rho}\exp\left({i}\int^xudx'\right){\phi}_2,\\
  & \frac{\prt{\phi}_2}{\prt x}=-i\sqrt{\rho}\exp\left(-{i}\int^xudx'\right){\phi}_1-\frac{i}{2}q{\phi}_2.
  \end{split}
\end{equation}
At last, we introduce the function
\begin{equation}\label{eq173}
  \psi=\sqrt{\rho}\exp\left({i}\int^xudx'\right),
\end{equation}
so Eqs.~(\ref{eq172}) become the Zakharov-Shabat spectral problem \cite{zs-73}
\begin{equation}\label{eq174}
\begin{split}
&  \left(
\begin{array}{c}
{\phi}_1 \\
{\phi}_2 \\
\end{array}
\right)_x = \mathbb{U}\left(
\begin{array}{c}
{\phi}_1 \\
{\phi}_2 \\
\end{array}
\right),\\
& \mathbb{U}=\left(
               \begin{array}{cc}
                 F & G \\
                 H & -F \\
               \end{array}
             \right)=\left(
               \begin{array}{cc}
                 \frac{i}{2}q & i\psi \\
                 -i\psi^* & -\frac{i}{2}q \\
               \end{array}
             \right).
\end{split}
\end{equation}
We suppose that the time-dependence of $\phi$-functions is governed by a similar linear system
\begin{equation}\label{eq175}
  \left(
\begin{array}{c}
{\phi}_1 \\
{\phi}_2 \\
\end{array}
\right)_t = \mathbb{V}\left(
\begin{array}{c}
{\phi}_1 \\
{\phi}_2 \\
\end{array}
\right),\quad \mathbb{V}=\left(
               \begin{array}{cc}
                 A & B \\
                 C & -A \\
               \end{array}
             \right).
\end{equation}
To find the elements of the matrix $\mathbb{V}$, we take again two basis solutions $(\phi_1,\phi_2)$
and $(\bar{\phi}_1,\bar{\phi}_2)$ of the system (\ref{eq174}), and build the squared basis functions
\begin{equation}\label{eq176}
  f=-\frac{i}{2}(\phi_1\bar{\phi}_2+\phi_1\bar{\phi}_1),\quad g=-\phi_1\bar{\phi}_1,
  \quad h=-\phi_2\bar{\phi}_2.
\end{equation}
The functions $\phi_1,\phi_2$ can be written in the form \cite{kamch-01}
\begin{equation}\label{eq177b}
  \begin{split}
  & \phi_1=\sqrt{g}\exp\left(i\int^2\frac{G}{g}dx'\right),\\
  & \phi_2=\sqrt{-h}\exp\left(i\int^2\frac{H}{h}dx'\right).
  \end{split}
\end{equation}
Now we take into account the identities \cite{kamch-94}
\begin{equation}\label{eq177}
  \left(\frac{G}{g}\right)_t-\left(\frac{B}{g}\right)_x=0,\quad
  \left(\frac{H}{h}\right)_t-\left(\frac{C}{h}\right)_x=0.
\end{equation}
In view of Eq.~(\ref{eq177b}), 
in quasiclassical approximation they must coincide with Eq.~(\ref{eq25}), so we get
\begin{equation}\label{eq178}
  B\approx-G\frac{\om}{k}\approx-\frac{i\psi}{2}(q+u)\approx-\frac{iq}{2}\psi-\frac12\psi_x,
\end{equation}
where we took into account the asymptotic relationship $u\approx-i\psi_x/\psi$, and a similar
formula can be written for $C$. The compatibility
condition $\mathbb{U}_t-\mathbb{V}_x+[\mathbb{U},\mathbb{V}]=0$ yields the system
\begin{equation}\label{eq179}
\begin{split}
  &F_t-A_x+CG-BH=0,\\ 
  &G_t-B_x+2(BF-AG)=0,\\ 
  &H_t-C_x+2(AH-CF)=0.
  \end{split}
\end{equation}
If we suppose that Eq.~(\ref{eq178}) is exact, that is $B=-\frac{iq}{2}\psi-\frac12\psi_x$,
and in a similar way $C=\frac{iq}{2}\psi^*-\frac12\psi_x^*$, then the first equation gives 
$A=a-\frac{i}{2}|\psi|^2$, where $a$ is constant. The other two Eqs.~(\ref{eq179}) yield 
$a=-\frac{i}{4}q^2$ and the compatibility condition reduces to
\begin{equation}\label{eq180}
  i\psi_t+\frac12\psi_{xx}-|\psi|^2\psi=0,
\end{equation}
provided
\begin{equation}\label{eq181}
  \mathbb{V}=\left(
               \begin{array}{cc}
                 -\frac{i}{4}q^2-\frac{i}{2}|\psi|^2 & -\frac{i}{2}q\psi-\frac12\psi_x \\ 
                 \frac{i}{2}q\psi^*-\frac12\psi_x^* & \frac{i}{4}q^2+\frac{i}{2}|\psi|^2 \\
               \end{array}
             \right).
\end{equation}
Thus, we have arrived at the celebrated Lax pair found by Zakharov and Shabat~\cite{zs-73} 
for the NLS equation which describes, in particular, the Gross-Pitaevskii dynamics of
Bose-Einstein condensate \cite{ps-03}.

\subsection{Derivative NLS equation}

As one more example, let us consider the derivative nonlinear  Schr\"{o}dinger (DNLS) equation
\begin{equation}\label{eq183}
  i\psi_t+\frac12\psi_{xx}-i(|\psi|^2\psi)_x=0,
\end{equation}
which finds applications, for example, to the theory of nonlinear Alfv\'{e}n waves (see, e.g.,
Ref.~\cite{kbhp-88}). At first, we should check if it satisfies the asymptotic integrability
condition. To this end, we transform it by means of the substitution (\ref{eq15})
to a hydrodynamic-like form
\begin{equation}\label{eq184}
  \begin{split}
  & \rho_t+\left[\rho\left(u-\frac32\rho\right)\right]_x=0,\\
  &u_t+uu_x-(\rho u)_x+\left(\frac{\rho_x^2}{8\rho^2}-\frac{\rho_{xx}}{4\rho}\right)_x=0.
  \end{split}
\end{equation}
Linearization of these equations yields the dispersion relation for harmonic waves
\begin{equation}\label{eq185}
  \om=k\left(u-2\rho\pm\sqrt{\rho(\rho-u)+\frac{k^2}{4}}\right),
\end{equation}
which corresponds to stable states for $\rho>u$. The dispersionless (hydrodynamic) limit
equations
\begin{equation}\label{eq186}
  \begin{split}
  & \rho_t+\left[\rho\left(u-\frac32\rho\right)\right]_x=0,\\
  &u_t+uu_x-(\rho u)_x=0
  \end{split}
\end{equation}
have real characteristic velocities
\begin{equation}\label{eq187}
  v_{\pm}=u-2\rho\pm\sqrt{\rho(\rho-u)}
\end{equation}
and can be transformed to a diagonal form (\ref{eq5}) by introduction of the Riemann invariants
\begin{equation}\label{eq188}
  r_{\pm}=\frac{u}{2}-\rho\pm\sqrt{\rho(\rho-u)}.
\end{equation}
As we see, the dispersion relation (\ref{eq185}), being expressed in terms of the Riemann
invariants (\ref{eq188}), takes the form of Eq.~(\ref{eq44}) with $\sigma=+1$. Hence, 
Eq.~(\ref{eq183}) is asymptotically integrable and we have the integral (see Eq.~(\ref{eq46}))
\begin{equation}\label{eq189}
\begin{split}
k^2 &=  (q+u-2\rho)^2-4\rho(\rho-u)\\
& =(q+u)^2-4q\rho
  \end{split}
\end{equation}
(we have redefined again the integration constant $q\to -q/2$ compared with Eq.~(\ref{eq46})). 
We consider this formula as a relationship obtained in a quasiclassical approximation 
applied to the system of first-order equations
\begin{equation}\label{eq190}
  \begin{split}
  & \frac{\prt\tilde{\phi}_1}{\prt x}=-\frac{i}{2}(q+u)\tilde{\phi}_1+i\sqrt{q\rho}\tilde{\phi}_2,\\
  & \frac{\prt\tilde{\phi}_2}{\prt x}=i\sqrt{q\rho}\tilde{\phi}_1+\frac{i}{2}(q+u)\tilde{\phi}_2,
  \end{split}
\end{equation}
so that the plane wave solution $\tilde{\phi}_1,\tilde{\phi}_2\propto \exp(ikx/2)$ satisfies the
condition (\ref{eq189}). Then the replacements (\ref{eq171}) and introduction of the function
(\ref{eq173}) coinciding with Eq.~(\ref{eq15}) yields the system
\begin{equation}\label{eq191}
  \begin{split}
  & \frac{\prt{\phi}_1}{\prt x}=-\frac{i}{2}q{\phi}_1+i\sqrt{q}{\phi}_2,\\
  & \frac{\prt{\phi}_2}{\prt x}=-i\sqrt{q}{\phi}_1+\frac{i}{2}q{\phi}_2.
  \end{split}
\end{equation}
To find the matrix $\mathbb{V}$ which determines the time dependence of $\phi$-functions according 
to Eq.~(\ref{eq175}), we use again formulas (\ref{eq177}) which yield in the quasiclassical 
approximation
\begin{equation}\label{eq192}
  \begin{split}
  & B=\frac{i}{2}q^{3/2}\psi-\frac12q^{1/2}\psi_x+iq^{1/2}|\psi|^2\psi,\\
  & C=-\frac{i}{2}q^{3/2}\psi^*-\frac12q^{1/2}\psi_x^*-iq^{1/2}|\psi|^2\psi^*,
  \end{split}
\end{equation}
and we assume that these expressions are exact. Then the first Eq.~(\ref{eq179}) gives
$A=a-\frac{i}{2}q|\psi|^2$, and the other equations reduce to Eq.~(\ref{eq183}) provided
$a=q^2/4$. Thus, we have arrived at the Lax pair in the matrix form (\ref{eq174}), (\ref{eq175}
with 
\begin{equation}\label{eq193}
\begin{split}
\mathbb{U}&=\left(
               \begin{array}{cc}
                 -\frac{i}{2}q & iq^{1/2}\psi \\
                 -iq^{1/2}\psi^* & \frac{i}{2}q \\
               \end{array}
             \right),\\
  \mathbb{V}&=\left(
  \begin{array}{cc}
  A & B \\
  C & -A
  \end{array}
             \right),
             \end{split}
\end{equation}
where
\begin{equation}\label{eq194}
  A=\frac{q^2}{4}-\frac{i}{2}q|\psi|^2
\end{equation}
and $B,C$ are defined in Eq.~(\ref{eq192}). It differs only by notation from the original form
that was found in Ref.~\cite{kn-78}.

As we see, quantization of a mechanical system associated with an integral of the
asymptotic integrability equations allowed us to derive in a very simple way the Lax pairs
of some typical completely integrable nonlinear wave equations.

\section{Conclusion}

We showed in this paper that the condition of asymptotic integrability of nonlinear wave 
equations leads to important relationships between the carrier wave number $k$ of 
high-frequency wave packets propagating along smooth background waves and the local 
values $(r_+,r_-)$ of the background variables. The existence of these relationships 
leads to many interesting applications. Besides immediate application to the theory of 
propagation of the high-frequency packets, it leads to the Hamiltonian theory of motion 
of narrow solitons along smooth, non-uniform, and time-dependent waves. Specification of 
this theory on the evolution of the edges of a dispersive shock wave allows one to 
calculate the number of solitons produced from an intensive wave pulse at asymptotically 
large time. The formula for the number of solitons turns out to be very general and it admits 
a general derivation as a consequence of the preservation of the Poincar\'{e}-Cartan 
integral invariant by the hydrodynamic background flow. This observation leads to a natural 
generalization in the form of the Bohr-Sommerfeld quantization rule for parameters of 
asymptotic solitons produced from the initial pulse. At last, the quasiclassical 
quantization according to the Bohr-Sommerfeld rule suggests that it follows from some full 
quantum theory of the corresponding mechanical system and the simplest Schr\"{o}dinger and 
Dirac recipes of finding such an underlying quantum theory yield the Lax pairs of 
the completely integrable equations. If the asymptotic integrability equations only have 
an approximate integral, then the Lax pair does not exist, but an approximate Bohr-Sommerfeld 
rule still can be written, so completely integrable equations share some of their properties 
with not completely integrable ones. It seems quite plausible that the theory of asymptotic 
integrability will lead to a number of other important consequences.

\section*{Acknowledgments}

I am grateful to E.~A.~Kuznetsov, M.~V.~Pavlov, S.~V.~Sazonov, D.~V.~Shaykin, and V.~V.~Sokolov for useful
discussions. This research is funded by the research project FFUU-2021-0003 of the Institute of Spectroscopy
of the Russian Academy of Sciences (Sections~1-3) and by the RSF grant number~19-72-30028
(Sections~4-7).

\end{document}